\documentclass[12pt,showpacs,preprintnumbers,floatfix]{article}
\usepackage{graphicx} 
\usepackage{dcolumn} 
\usepackage{bm} 
\usepackage{amsfonts}
\usepackage{amsthm}
\usepackage{amsmath}
\usepackage{amssymb}
\usepackage{epsfig}
\usepackage{color}
\usepackage{textcomp}
\usepackage{hyperref}
\usepackage{titlesec}
\usepackage{slashed}
\usepackage{caption}
\usepackage{subcaption}

\def\be{\begin{equation}}
\def\ee{\end{equation}}
\def\ba{\begin{eqnarray}}
\def\ea{\end{eqnarray}}

\def\bdm{\begin{displaymath}}
\def\edm{\end{displaymath}}

\def\bq{\begin{quote}}
\def\eq{\end{quote}}

 at 10truept

\newcommand{\bea}{\begin{eqnarray}}
\newcommand{\eea}{\end{eqnarray}}

\newcommand{\bi}{\begin{itemize}}
\newcommand{\ei}{\end{itemize}}

\newcommand{\beq}{\begin{equation}}
\newcommand{\eeq}{\end{equation}}
\newcommand{\beqa}{\begin{eqnarray}}
\newcommand{\eeqa}{\end{eqnarray}}

\newcommand{\nn}{\nonumber}

\def\ltap{\ \raise.3ex\hbox{$<$\kern-.75em\lower1ex\hbox{$\sim$}}\ }
\def\gtap{\ \raise.3ex\hbox{$>$\kern-.75em\lower1ex\hbox{$\sim$}}\ }
\def\gl{\ \raise.5ex\hbox{$>$}\kern-.8em\lower.5ex\hbox{$<$}\ }
\def\roughly#1{\raise.3ex\hbox{$#1$\kern-.75em\lower1ex\hbox{$\sim$}}}

\begin{document}

\thispagestyle{empty}
\begin{flushright}
September 2017\\
TUM-HEP 1101/17
\end{flushright}
\vspace*{1.0cm}
\begin{center}
{\Large \bf Theory and Phenomenology of\\ Planckian Interacting Massive Particles as Dark Matter}\\

\vspace*{1.0cm} {\large Mathias Garny$^{a,}$\footnote{\tt
mathias.garny@tum.de}, Andrea Palessandro$^{b,}$\footnote{\tt
palessandro@cp3.sdu.dk},\\ McCullen Sandora$^{c,}$\footnote{\tt
mccullen.sandora@tufts.edu},  Martin S. Sloth$^{b,}$\footnote{\tt
sloth@cp3.sdu.dk}}\\
\vspace{.5cm} {\em $^a$Physik Department T31, Technische Universit\"at M\"unchen,
James-Franck-Stra\ss e 1, D-85748 Garching, Germany}\\
\vspace{.5cm} {\em $^b$CP$^3$-Origins, Center for Cosmology and Particle Physics Phenomenology \\ University of Southern Denmark, Campusvej 55, 5230 Odense M, Denmark}\\
\vspace{.5cm} {\em  $^c$Institute of Cosmology, Department of Physics and Astronomy\\ Tufts University, Medford, MA 02155, USA}

\end{center}

\begin{abstract}Planckian Interacting Dark Matter (PIDM) is a minimal scenario of dark matter assuming only gravitational interactions with the standard model and with only one free parameter, the PIDM mass. PIDM can be successfully produced by gravitational scattering in the thermal plasma of the Standard Model sector after inflation in the PIDM mass range from TeV up to the GUT scale, if the reheating temperature is sufficiently high. The minimal assumption of a GUT scale PIDM mass can be tested in the future by measurements of the primordial tensor-to-scalar ratio. While large primordial tensor modes would be in tension with the QCD axion as dark matter in a large mass range, it would favour the PIDM as a minimal alternative to WIMPs.  Here we generalise the previously studied scalar PIDM scenario to the case of fermion, vector and tensor PIDM scenarios, and show that the phenomenology is nearly identical, independent of the spin of the PIDM. We also consider the specific realisation of the PIDM as the Kaluza-Klein excitation of the graviton in orbifold compactifications of string theory, as well as in models of monodromy inflation and in Higgs inflation. Finally we discuss the possibility of indirect detection of PIDM through non-perturbative decay.
\end{abstract}

\vfill \setcounter{page}{0} \setcounter{footnote}{0}
\newpage

\tableofcontents

\section{Introduction}

For at least the better part of a century now, the general belief has been that we are just the scum of the universe.  The bulk is dark matter, consisting of particles that do not interact electromagnetically.  In a sense we are lucky that we know about it at all, as the equivalence principle guarantees that it interacts with us at least gravitationally, giving us a window into an otherwise inaccessible realm.  Although physicists are usually hopeful that dark matter couples to the standard particle content of the universe through some additional force that will source dramatic new signatures, the extent of our knowledge at the moment only requires this gravitational coupling.  In the past, there has been good reason to suspect that some additional coupling may be present: the WIMP miracle attests that weak-scale dark matter happens to produce the observed relic abundance, and many scenarios of beyond the standard model physics include extra particles at this scale, some of which may be potential dark matter candidates.

And yet, even after decades of search, this paradigm remains unverified.  Direct detection experiments have produced a distinct lack of signals, and the evidence for new physics at the TeV scale has not been forthcoming from the LHC.  While it is premature to claim the WIMP paradigm is effectively ruled out, its initial seductiveness has been somewhat diminished, as the originally expected values of its parameters are increasingly pushed toward heavier masses and weaker couplings.

\subsection{A GUT scale coincidence versus additional new physics} 

If the WIMP paradigm turns out to be a blind alley, then either the solution to the hierarchy problem simply does not entail a dark matter candidate, as in low-scale SUSY with broken $R$-parity, or instead we are being coerced toward a fundamentally different view of nature: that the particle content of the standard model is all that there is, unaccompanied by the panoply of light states that would conspire to render the weak scale technically natural. In this potentially impending realization, the standard model is simply not typical, as far as the set of possible laws of physics is concerned.  The fact that it occurs 15 orders below the Planck scale, and appears to require a wild fine-tuning of the fundamental parameters in order to maintain this hierarchy, may be reality.  The one reason that we might accept such an egregious setup, however, is that the presence of life may actually be contingent on this miraculous conspiracy.  The weak scale happens to be very close to the QCD scale, and as a consequence many different atoms exist, and can exchange energy over extremely long timescales, that sets up the existence of building blocks for life, as well as stable environments where it can thrive \cite{Susskind:2003kw,Agrawal:1998xa,Feldstein:2006ce,Donoghue:2009me}.

A milder anthropic argument requires that the dark matter abundance be not too much larger or smaller than its observed value \cite{Linde:1987bx,Turner:1990uz,Linde:1991km,Garriga:2002tq,Garriga:2003hj, Wilczek:2004cr,Hellerman:2005yi,Tegmark:2005dy,Hertzberg:2008wr,Freivogel:2008qc}, otherwise the structure needed for our existence would not have formed.  Aside from the abundance, however, any other details of the dark sector are not crucial for life, and so one would not expect them to be as fine tuned as the standard model. Even if we are willing to accept an anthropic argument for the unnaturalness of the weak scale, we would expect the dark sector to be as natural as possible, a generic instantiation of physical law.  If this is the case, the particle spectrum would be much closer to the Planck scale, the natural ultraviolet cutoff scale.  

Indeed, we expect new physics to come in close to the Planck scale, as at these energies gravitational scattering violates perturbative unitarity bounds.  A commonly referenced scale is perhaps two orders of magnitude below, which is typically the size of the extra dimensions in perturbative string theory, and where the fundamental forces of natural appear to (almost) unify into a single GUT theory.  Two more orders of magnitude below that, sterile neutrinos may reside, giving rise to a see-saw explanation of the tiny neutrino masses we observe. Another indication of new physics near the GUT scale is the conventional viewpoint that primordial inflation is needed to solve the flatness and the horizon problem. Large field models of inflation rooted in string theory require inflation to happen at an energy scale just below the GUT scale in order for the amplitude of the perturbations to match the CMB observations. If the dark matter is part of this realm, it would naturally fit in with known completions of particle physics and gravity, and reinforce exactly how special the laws of our visible sector are.

The immediate problem with such an idea is that it is extremely hard to produce such heavy states.  If our philosophy is taken literally, then the dark matter does not even have to couple to the inflationary sector (or whatever field reheated the universe), and so is not produced directly in the early universe. WIMPs start in thermal equilibrium, then gradually decouple from the plasma as the density becomes too low for them to continue to efficiently interact.  Particles that are only gravitationally coupled are never in equilibrium with the dominant plasma of the universe, but nevertheless they may be produced with the right abundance via the freeze-in mechanism \cite{Hall:2009bx}, which relies on annihilations of the standard model particles to populate the dark sector.  This is a relatively rare process even when the dark matter mass, $m_X$, is not Planckian, and it becomes all the more difficult to bear out if it does. This PIDM production mechanism can successfully produce the right dark matter abundance in the large PIDM mass range \cite{Garny:2015sjg}
\beq
1~ \textrm{TeV} \lesssim m_X \lesssim M_{\textrm{GUT}}
\eeq
if the reheating temperature is sufficiently high. Note, that since we are either assuming that either $R$-parity is broken, or there is no low-scale supersymmetry, there is no lightest supersymmetric partner (LSP), and therefore the usual upper bound on the reheating temperature $T_{rh}\lesssim 10^{10}$GeV, from the LSP not overclosing the universe, is not valid. 

In fact, as we show in section (\ref{AbC}), obtaining the right abundance of dark matter in the heavy PIDM limit, $m_X > 2.5\times 10^{-6}\gamma^{-7/8}m_p$ ($\gamma=1$ for instantaneous reheating), determines up to logarithmic corrections the reheating temperature
\beq
T_{rh} \simeq \sqrt{\gamma} m_X ~.
\eeq
We see that with a reheating temperature just below the GUT scale, a PIDM with a GUT scale mass can lead to the right dark matter abundance. We note, however, that in the heavy PIDM limit, the dark matter abundance is very sensitive to $m_X$ for a fixed $T_{rh}$, while in the low mass regime the abundance scales only linearly with $m_X$ for fixed $T_{rh}$.

In the PIDM scenario, dark matter is as decoupled as fundamentally allowed, making it the hardest scenario of dark matter to possibly test, and all hopes of direct detection, indirect detection, and collider searches are doomed to fail.  However, the inefficiency of the production mechanism offers a silver lining, as the reheating temperature must be close to the GUT scale if adapting minimal assumptions of a GUT scale PIDM mass.  The only way such a setting can be arranged is if both the efficiency of reheating and the energy scale of inflation are very high. This, in turn, implies that primordial tensor modes will be necessarily produced alongside the PIDM.  If tensor modes are not detected in the next round of proposed experiments, the interesting parameter regime of GUT scale masses will be definitively ruled out, and would be an indication that there is either some additional interaction between the dark matter and the visible sector or some additional mass scale present in the universe.

\subsection{Outline}
We have demonstrated the ability of the PIDM scenario in our previous work, \cite{Garny:2015sjg}.  In section \ref{calcset}, we elaborate on this initial investigation, and then in the following sections we present two crucial features that were not addressed in the first work: in section \ref{crosscalcs}, we systematically run through the setup for various different choices for the spin of the PIDM, and show that the outcome is extremely insensitive to this feature of the dark matter.  
The second feature we emphasize in section \ref{insitu} is that the PIDM paradigm, which before was treated as a phenomenological Lagrangian, can be incorporated into existing models of high-energy physics.  We demonstrate this in several models, starting with a Kaluza-Klein setup, then moving on to monodromy models of inflation.  In both of these, the degrees of freedom already known to be present can naturally act as dark matter through this mechanism, and no additional physics need be invoked for this scenario to be borne out.  We end the section with a discussion of how the PIDM fits into Higgs inflation, a minimalistic model of the inflationary sector. 
Finally, in section \ref{posssig} we elaborate on one potential additional signature of the PIDM that was briefly mentioned in \cite{Garny:2015sjg}, namely PIDM decay mediated by gravitational instantons.  This produces ultra high energy cosmic rays that may be detectable in future observatories.  Because the bounds on these processes are already extremely strong, the action suppressing this decay channel must be $\mathcal{O}(100)$, quite a bit larger than that found in standard Einstein gravity calculations, but easily accommodated in generic quantum gravity extensions. We also briefly discuss the possible decay through perturbative gravitational portals, which are absent in flat space. If the signal does turn out to be measurable, it will provide a rare window with which the nature of quantum gravity may be probed.

Throughout this paper  $m_p=1.2\times 10^{19} \text{GeV}$ is the non-reduced Planck mass.

\section{Calculational Setup}\label{calcset}

In this section we calculate the production of dark matter assuming that dark matter only couples to the standard model through gravity.

\subsection{Production}
  The Lagrangian for such a scenario will simply be
\beq
{\cal L} = {\cal L}_{SM} + {\cal L}_{DM} + {\cal L}_{EH}+\frac{\sqrt{8\pi}}{2m_p}h^{\mu\nu}(T^{SM}_{\mu\nu}+T^{DM}_{\mu\nu}) \,.
\eeq
Here, ${\cal L}_{SM}$ denotes the SM Lagrangian, ${\cal L}_{DM}$ governs the dark matter, and ${\cal L}_{EH}$ is the Einstein-Hilbert Lagrangian. We are assuming that some global charge is preventing the direct coupling between the PIDM and SM, and preventing the PIDM from decaying. As an example, in section (\ref{insitu}) we will discuss the concrete model where the PIDM is the Kaluza-Klein excitation of the graviton, and Kaluza-Klein parity conservation protects it from decaying. It has been conjectured, however, that in a full UV complete theory of quantum gravity, there are no global symmetries \cite{Banks:2010zn}, and so we return to the interesting possibility of an observable PIDM decay signal induced by quantum gravity in section (\ref{posssig}).

\begin{figure}[t]
\begin{centering}
\includegraphics[width=6cm]{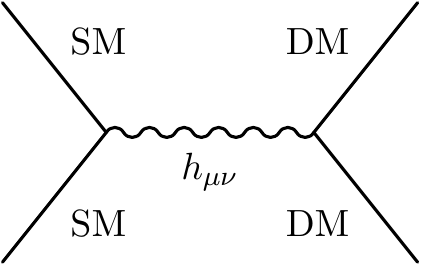}
\caption{The main contribution to PIDM production comes from annihilation of standard model particles into a graviton and subsequent pair production of the dark matter particles.}
\label{freeze}
\end{centering}
\end{figure}

For simplicity, dark matter here is taken to be a free particle with no self interactions, though in principle some could be added.  Communication between the two sectors is only mediated indirectly through gravity, which couples to their energy-momentum tensors.  Because this coupling is fixed by the equivalence principle, the only free parameter in this model is the mass of the PIDM.  If this were to be written as an effective interaction in the Lagrangian by integrating out the graviton it would be a dimension 8 operator for a scalar PIDM, albeit a nonlocal one since the graviton is massless. The annihilation of standard model particles into dark matter that comes from this Lagrangian is through s-channel graviton exchange depicted in Fig. \ref{freeze}, and results in a $2\rightarrow 2$ amplitude

\beq\label{MatrixElement}
 {\cal M} = \frac{-i 8\pi\langle p_1 |T_{SM}^{\mu\nu}|p_2\rangle\langle k_a|(T^{DM}_{\mu\nu}-\frac12g_{\mu\nu}T^{DM\alpha}_{\phantom{DM} \alpha})|k_b\rangle}{m_{p}^2(k_a+k_b)^2}\,.
\eeq
Here, diffeomorphism invariance forces the on-shell energy momentum tensors to be divergence free, $(k_a+k_b)^\mu T_{\mu\nu}=0$.

This annihilation must have been crucial in the early universe for dark matter to be produced at all, since we have dictated that it does not even couple to the inflaton sector.

Since the interactions between the Standard Model and the hidden sector are Planck suppressed, we expect them to be
effective only at the highest available energies. Indeed, there are a number of well-known mechanisms to produce a
population of either super-heavy or super-weakly interacting dark matter particles.  In this work we rely on the freeze-in method for production, by which a hot bath of high temperature standard model particles occasionally results in collisions that annihilate and pair produce dark matter.  This is a minimal scenario because this process automatically takes place, and contributions from other production mechanisms will be commented on in section \ref{otherprod}.

\begin{table}[h]
\vskip.4cm
\begin{center}
\begin{tabular}{|c|c|c|c|}
\hline
\multicolumn{4}{|l|}{Beyond WIMP}\\
\hline  & Super Light & Light & Heavy \\
\hline
Strong & x & x &  MACHOs, black holes\\
\hline
 Weak & x & WIMP &  WIMPZILLA\\
 \hline
 Super Weak   &  Axion, sterile $\nu$ &  gravitino, FIMP & PIDM\\
\hline
\end{tabular}
\end{center}
\caption{Summary of various dark matter proposals, organized according to how strongly they interact with the SM and how heavy the particles are. 
}
\label{tableone}
\end{table}

\subsection{Relation to Other Dark Matter Scenarios}

Freeze-in production of super-weakly interacting dark matter particles has been studied in many different contexts, mostly focusing on light (electro-weak scale) masses (\cite{McDonald:2001vt,Choi:2005vq,Kusenko:2006rh,Petraki:2007gq,Hall:2009bx}). These particles never reach thermal equilibrium, but any particle that is produced via its (rare) interactions with the SM accumulates and contributes to its final abundance.
It is important to discriminate whether the interactions are mediated by renormalizable or non-renormalizable interactions.
In the former case, the production is dominated by the lowest temperatures $T \sim m_X$ for which it is kinematically allowed, while in the latter
case the production is dominated by the highest available temperatures.

In the present case, we are concerned with a scenario where the dark matter is {\it both} super-heavy and super-weakly interacting,
mediated by its (non-renormalizable) gravitational interaction. We therefore expect it to be dominated by the highest available energy scales
after the end of inflation, encompassing both a non-thermal contribution from production during re/preheating, and a thermal contribution
from freeze-in during radiation domination. The former is dominant for heavy dark matter with mass at or above the maximal temperature
after inflation $m_X\gtrsim T_{\rm MAX}$ (see \eqref{maxtem} for an estimate).

Note the PIDM scenario differs in important ways from its cousin, the WIMPZILLA \cite{Kolb:1998ki,Chung:2001cb}. In one version of the WIMPZILLA, the WIMPZILLA is produced by the change in the vacuum at the end of inflation. This production mechanism, which the authors labelled gravitational production, is only effective for a narrow mass range well below the GUT scale, but in this restricted mass range the PIDM becomes similar to a WIMPZILLA (see also section (\ref{otherprod})).  In another version of  the WIMPZILLA the authors also considered a different production mechanism where dark matter is assumed to be coupled to the standard model via gauge interactions, like the WIMP but much heavier. Thus, in this version of the WIMPZILLA the thermally averaged cross-section is assumed to have the form $\left<\sigma v\right> \sim \alpha_X /m_X^2$, where $\alpha_X$ is the gauge coupling strength of the WIMPZILLA to the standard model, and $m_X$ is the mass of the WIMPZILLA. Therefore this production mechanism is also different than the production mechanism for the PIDM.  

A related philosophy to the PIDM has also been followed in some works on hidden charged dark matter\footnote{For some more recent and different directions of exploration related to the PIDM, see also \cite{Tang:2016vch,Ema:2016hlw,Kannike:2016jfs,Babichev:2016bxi,Bernal:2017kxu,Baym:2017xvh}.} \cite{Feng:2009mn,Agrawal:2016quu}. Here it is sometimes assumed that the dark matter sector is decoupled completely from the standard model and produced by freeze out in the dark sector only. Dark matter can then only be probed by the imprint on large scale structure of its self-interactions mediated by dark photons. In these models, it is typically not specified how the dark sector is reheated after inflation. It would be interesting to use the production mechanism of the PIDM to explore PIDMs carrying a $U(1)$ charge in the dark sector, since PIDMs with a mass as small as $10^3$\,GeV in principle are allowed \cite{Garny:2015sjg}. We leave this for future work, and explore the minimal PIDM scenario here.

\subsection{Boltzmann equations}

PIDM production occurs dominantly at the highest energy scales after inflation, namely in the periods shortly after reheating, during reheating, and at the end of inflation.  We find that generically the reheating contribution dominates, so to compute the abundance we use the description of the reheating dynamics from \cite{Giudice:2000ex}, 

\bea
  \frac{d\rho_\phi}{dt} &=& -3H(1+w)\rho_\phi- S \label{phievol}\\
  \frac{d\rho_R}{dt} &=& -4H\rho_R + S +2\langle\sigma v\rangle \langle E_X\rangle \left(n_X^2-(n_X^{eq})^2\right)\label{radevol}\\
  \frac{d n_X}{dt} &=& -3H n_X - \langle\sigma v\rangle\left(n_X^2-(n_X^{eq})^2\right)\label{pimpevol}
\eea
These dictate how the energy density of the inflaton (or, more generally, reheaton) $\rho_\phi$, radiation $\rho_R$ and number density of dark matter $n_X$ evolve.  These sectors communicate through the term $S$, which describes the inflaton decay into relativistic Standard Model, and which encapsulates all dependence on the reheating dynamics. The effective equation of state $w$ captures the dynamics of the inflaton field.

For a general preheating scenario both $S$ and $w$ can have a complicated time dependence. Here we make the minimal assumption that $S=\Gamma \rho_\phi$ where both $\Gamma$ and $w$ are constant. If we denote the Hubble rate at the end of inflation by $H_i$, reheating will be practically instantaneous (within one Hubble time) for $\Gamma\to H_i$, while $\Gamma\ll H_i$ gives rise to perturbative reheating. The reheating temperature, defined by the condition $H=\Gamma$ is

\beq
T_{rh}=\kappa_2\gamma(m_pH_i)^{1/2}\label{trh}\,.
\eeq

Here $\kappa_2=(45/(4\pi^3g_{rh}))^{1/4}\approx.25$, and $g_{rh}$ is the number of degrees of freedom at reheating, which we will assume to be that of the SM, though, since it enters with the one-fourth power, our conclusions are quite insensitive to the precise number of particles.  This defines the constant $\gamma\in(0,1)$, which parameterizes the efficiency of reheating.
A value of $\gamma=1$ corresponds to perfectly instantaneous reheating.  It can be expressed as 
\beq
\gamma=\sqrt{\frac{\Gamma}{H_i}}=\left(\frac{g_{i}}{g_{RH}}\right)^{1/4}e^{-\frac34N_{RH}(1+w)}
\eeq
where $N_{RH}$ is the total duration of the reheating process, in e-folds.  In order for the freeze-in scenario to work in this context, $\gamma$ must be close to 1.

Such an efficient scenario has been achieved in several concrete realizations, even in a perturbative reheating setup.  For this, the decay rate is $\Gamma=g^2m_{\phi}/(8\pi)$, where $g$ is the coupling between the inflaton and the SM, and so $\gamma\approx .2g\sqrt{m_\phi/H_i}$.  This can easily be large enough to produce the PIDM, if the inflaton is not too weakly coupled to the radiation and the mass of the inflaton during reheating is comparable to the Hubble rate.  However, one might worry that such a strong coupling could lead to a naturalness problem in the inflaton sector \cite{Kofman:1996mv}, as standard model degrees of freedom running through loops will alter the flatness of the inflaton potential.  This might motivate a generalization of the present study to include some specific cases of non-perturbative reheating, which can be very efficient, though heavily dependent on particular details of the model at hand.  For instance, a non-perturbative scenario capable to achieving $\gamma\approx1$ was laid out in \cite{Felder:1998vq}, though the mechanism they find should not be expected to be a generic feature of inflationary models. Here, we restrict our attention to perturbative scenarios, since the results are more generic, and less model-dependent. In section \ref{mono}, we will discuss an example with perturbative reheating where symmetries forbid radiative corrections from destroying the flatness of the inflaton potential even if $\gamma=1$. This is achieved in a variant of natural inflation (monodromy inflation), where a broken shift symmetry protects the flatness of the inflaton potential.

Now we turn to the evolution of the system of equations (\ref{phievol}-\ref{pimpevol}).  Assuming that the energy density in the inflaton field dominates in equation (\ref{phievol}), it will dilute as 
\beq
\rho_\phi\propto a^{-3(1+w)}
\eeq
until the point where $H=\Gamma$ and the inflaton decays, signalling the end of reheating.

The radiation density, which can be related to the temperature of the plasma through $\rho=g\pi^2T^4/30$, has a complicated time dependence, as the initial source of energy heats the plasma, becomes depleted and the entire system starts to cool adiabatically.

This provides the relation between temperature and scale factor needed in order to calculate the abundance.  In the reheating phase the temperature behaves as~\cite{Giudice:2000ex}
 \beq
 T(a)=\frac{\kappa_1(\gamma m_p H_i)^{1/2}}{(1+3/5w)^{1/4}}\left(a^{-3(1-w)/2}-a^{-4}\right)^{1/4}\,,\label{tempa}
 \eeq
where $\kappa_1=(9/(2\pi^3g_{max}))^{1/4}\approx .20$, 

The temperature attains a maximum at the value of the scale factor  
\beq
a_{max}=\left(\frac{8}{3(1-w)}\right)^{2/(5+3w)}\,,
\eeq
 in the convention where $a_i=1$ at the beginning of reheating.  At this time, the temperature is 
\beq\label{maxtem}
T_{max}=\kappa_1\left(\frac58\right)^{1/4}\left(\frac{8}{3(1-w)}\right)^{\frac{3(1+w)}{4(5+3w)}}(\gamma m_p H_i)^{1/2}\,.
\eeq
The dependence on the equation of state is relatively mild unless $w\approx1$, but the maximum temperature scales like the square root of the efficiency parameter $\gamma$, so is roughly the geometric mean of the actual and instantaneous reheat temperatures.

After reheating the temperature evolves as
\beq
T(a)=T_{rh}\frac{a_{rh}}{a}\;,
\eeq
and the Hubble rate is
\beq\label{hofa}
  H \simeq H_i \left\{ \begin{array}{ll} (a/a_i)^{-3(1+w)/2} & a<a_{rh} \\
  \gamma^2 (a/a_{rh})^{-2} & a>a_{rh} \end{array}\right.\,,
\eeq
with $a_i$ being the scale factor at the end of inflation and $a_{rh}=a_i\gamma^{-\frac{4}{3(1+\omega)}}$ the scale factor at the end of reheating.

The last term in equation (\ref{radevol}) is the collision contribution, where $\langle\sigma v\rangle$ is the usual thermally averaged $2\to 2$ cross section for dark matter pair annihilation into SM particles.  The precise form of this depends on the details of the PIDM spin, and is calculated for each case in section \ref{crosscalcs}. Because the PIDM is so weakly interacting, it will never come to dominate the evolution of the plasma.

To solve the system \ref{pimpevol} we will need the equilibrium value $n_{eq}$, which is given by  \cite{Giudice:2000ex}
\beq
n_X^{eq}=\frac{g}{2\pi^2}m_X^2TK_2\left(\frac{m}{T}\right)
\eeq
where $K_2(x)$ is the modified Bessel function.

We also attempt to address the issue that we have been placing bounds on $H_i$, the Hubble rate at the onset of reheating, and trying to relate this to $H_{CMB}$, the scale of inflation as observed in the cosmic microwave background.  In general, $H_i<H_{CMB}$.  However, we can do slightly better, to estimate the magnitude of the deviation during the ensuing evolution.  Using $dH/dN=-\epsilon H$, 
\beq
\Delta H=-\int_{N_{CMB}}^{N_{rh}}dN\epsilon H\approx-.0375\left(\frac{r}{.01}\right)\left(\frac{\Delta N}{60}\right)H
\eeq
where we have made the assumption that $\epsilon H$ is nearly constant during evolution, and used the single field slow roll result $r=16\epsilon$.  This manifestly assumes that inflation does not end by $\epsilon$ becoming large, but rather the other slow roll parameter, $\eta$, which is expected in many small field models of inflation\footnote{In large field models of inflation with a monomial inflaton potential of the form $V(\phi)\propto \phi^\alpha$, which ends by $\epsilon$ becoming large, one finds $H_{CMB}/H_i \sim N_{CMB}^{\alpha/4}$, which is less than an order of magnitude for $\alpha\lesssim 2$.}. Under these approximations, we see that the order of magnitude change in $H$ from CMB to reheating can be considerable for current bounds on $r$, but if it is constrained by another order of magnitude or so, the difference in energy scales will become negligible. 

Foregoing this estimate, however, the combined CMB bound on tensor modes, $r<0.07$ (95\% CL) \cite{Array:2015xqh, Ade:2015lrj}, implies an upper bound on $H_i$,
\beq
H_i<6.6\times10^{-6}m_{p} \left(\frac{r}{0.1}\right)^{1/2}\;.
\eeq  

\subsection{Abundance Calculation}\label{AbC}

With the previous assumptions, the energy density of the inflaton and radiation can be solved, leaving the quantity of interest for our purposes, the number density of dark matter particles.  Because it interacts so weakly, the dark matter will never come to dominate the energy density, or even reach equilibrium, and the differential equation can be simplified in that case.  It is convenient to define the dimensionless abundance $X=n_X a^3/T_{rh}^3$, which obeys
\beq\label{Xeq}
\frac{dX}{da}=\frac{a^2}{T_{rh}^3H(a)}\langle\sigma v\rangle (n_X^{eq})^2\,.
\eeq
This is valid when $n_X \ll n_{eq}$, and the inverse annihilation process dominates. 

  One interesting feature of this equation is that it is valid independent of the momentum distribution of the dark matter phase space density. This means that we do not have to make the assumption that the dark matter is thermally distributed, which is relevant in this context, because PIDMs are so weakly (self-)interacting that once they are produced they likely will never interact again.  Even so, this equation will hold for the zeroth moment of the phase space distribution, namely the number density.
  
 Note that we have used Maxwell-Boltzmann statistics to derive these expressions, which is manifestly a good approximation in the large mass limit, where the exponential suppression keeps population levels low enough that quantum effects are not important.  This is less obviously valid in the high temperature limit, which is naively in the opposite regime.  However it turns out to be a fine approximation in this regime as well, since the process is given by a nonrenormalizable interaction, leading to the scaling $|{\cal M}|^2\sim E^4/m_p^4$.  Because of this, the highest momenta, $p\sim T$, which are not Boltzmann suppressed, dominate the phase space integration and corrections from quantum statistics are therefore not large.

The equation for $X$ can be integrated directly, assuming the initial abundance vanishes
\beq\label{Xfint}
X_f=\frac{1}{T_{rh}^3}\int_{a_i}^{a_f}da\frac{a^2}{H(a)}\langle\sigma v\rangle(n_X^{eq})^2\;.
\eeq
To perform this integral, we use (\ref{trh}), (\ref{hofa}), (\ref{sigv}).  We are able to send $a_f\rightarrow\infty$ without introducing any noticeable error, since the production rate is exponentially suppressed by this point, making the integral insensitive to this region.  Once this integration has been performed it will yield an expression for the number density in terms of four parameters, $m_X,H_i,\gamma$, and $w$.  This is related to the present day relic abundance through
\beq\label{Omhsq}
\Omega_X h^2=Q\gamma^{\frac{4}{1+w}}\frac{m_X}{m_p}X_f,\quad Q=\frac18\,\frac{T_{rh}^3m_ps_0}{s_{rh}\rho_c} \approx 9.2\times10^{24}\,.
\eeq
Here $\rho_c=1.88\cdot 10^{-29}$g/cm$^3$ is the critical density for $h=1$ and $s_0(s_{rh})$ is the present entropy density today (at $T=T_{rh}$).
The factor $1/8$ is as introduced in \cite{Giudice:2000ex}, parametrizing the fact that entropy continues to be generated for a time after reheating.

This integral is done numerically for several varieties of PIDM in the following section, but it is useful to calculate it analytically in the computationally tractable ``heavy PIDM limit", $m_X\gg T$.  Here the equilibrium density asymptotes to an exponential.  We work with a scalar PIDM for definiteness, and set $w=0$, as it only complicates matters without adding to the physical intuition we are trying to build.  The abundance becomes
\beq
X_f=\frac{N_0m_X^5}{8\pi^2m_p^4T_{rh}^3H_i}\left[T_1^3\int_1^{a_{rh}}daa^{19/8}e^{-\frac{2m_X}{T_1}a^{3/8}}+\frac{a_{rh}}{\gamma^2}T_{rh}^3\int_{a_{rh}}^{a_{f}}da\frac1ae^{-\frac{2m_X}{T_{rh}}\frac{a}{a_{rh}}}\right]\;.
\eeq
Here, $N_0$ is the number of scalar particles in the standard model, the effects of the other degrees of freedom are subdominant, and $T_1=(1728/3125)^{1/20}T_{max}=.97 T_{max}$.  We have split the integral into thermal and nonthermal contributions, and treated the thermal contribution during reheating to occur after the maximal temperature in (\ref{tempa}) has already been reached.  In the limit where $a_{rh}\gg1$, $a_f\gg a_{rh}$, these integrals can be done, yielding
\beq
X_f=\frac{N_0m_X^4}{8\pi^2m_p^4T_{rh}^3H_i}\left[\frac43T_1^4e^{-\frac{2m_X}{T_1}}+\frac12\frac{a_{rh}}{\gamma^2}T_{rh}^4e^{-\frac{2m_X}{T_{rh}}}\right]\;.
\eeq
For $\gamma\ll1$, the maximum temperature will be much larger than the reheating temperature, and the second term can be neglected.  The resultant expression can be set equal to the observed relic abundance and solved for $H_i(m_X)$.  The expression is:
\beq
H_i(m)=\frac{4m_X^2}{\kappa_1^2\gamma m_p}W_{-1}\left(-\lambda\gamma^{-7/2}\frac{m_p^4}{m_X^4}\right)^{-2}\;.
\eeq
Here $W_{-1}(x)$ is the Lambert productlog function, and $\lambda=12\pi^2\Omega_Xh^2\kappa_2^3/(\kappa_1^4Q)=1.4\times10^{-23}$.  This expression gives us some insight into the basic properties of the Hubble rate needed in order to produce the correct dark matter abundance:  Firstly, if the argument of the productlog is less than $-1/e$, no real solution exists.  This in turn places a restriction on the parameters of the model: $\gamma^{7/8}m_X>2.5\times10^{-6}m_p$.  For small masses the heavy PIDM approximation breaks down, invalidating the analysis we used, but for large masses, this gives a lower bound on the efficiency of reheating necessary for this production mechanism to work.

When the mass is large, the productlog is well approximated by a logarithm, and so in this regime the Hubble rate necessary to produce the correct dark matter abundance scales as $H_i\propto m^2/(\gamma\log(\gamma^{7/8}m_X)^2)$.  This scaling is borne out by the exact results presented in section \ref{crosscalcs}.

\subsection{Relation to Other Production Mechanisms}\label{otherprod}

Before turning to the full calculations, we briefly contrast the freeze-in mechanism to other potential sources of PIDM dark matter.  These end up being not as relevant in the most interesting parameter regime, where the dark matter mass is near the Planck scale.

The first additional method of production is due to the time-dependence of the metric at the end of inflation, which triggers a Bogoliubov-type creation of particles.  This mechanism has been called `gravitational production' \cite{Chung:2001cb,Kuzmin:1999zk}.  It is most relevant for $m_X \simeq H_i$, which is away from the heavy PIDM limit, and leads to modifications only within a relatively narrow range of masses, around $m_X\sim 10^{-10}-10^{-7}m_p$, depending on the efficiency of reheating.  For this contribution to the energy density, we use the result from \cite{Chung:2004nh}.

An additional mechanism can occur for the case of scalar PIDMs in the low mass regime $m_X \ll H_i$: during inflation, these act as a field, and so locally acquire a vacuum expectation value typically of the order $X \sim \sqrt{\left< X^2\right>} = H_i/2\pi$.  This condensate will actually be the dominant energy density of the PIDM (unless the potential of the PIDM is modified away from the pure quadratic form for large field values, as in the axion potentials, for example), and lead to DM isocurvature perturbations, that are already ruled out observationally \cite{Ade:2015lrj} (see \cite{Hertzberg:2008wr,Seckel:1985tj,Nurmi:2015ema} for related discussions).  So, this regime is not the most enticing part of parameter space for these models, and its presence is highly model dependent.

\section{Specific PIDM Calculations}\label{crosscalcs}
Up to this point we have left the calculation of the relic abundance general, in terms of a cross section for freeze-in.  The precise value of this will depend on the type of particle that the PIDM is.  In this section we compute the cross section for several different choices of spin.  Generically, PIDM production will be a sum of three contributions:
\beq
\langle\sigma v\rangle=N_0 \langle\sigma v\rangle_0+N_{1/2}\langle\sigma v\rangle_{1/2}+N_1\langle\sigma v\rangle_1\,,
\eeq
where the subscripts denote the spin of the standard model (SM) particles, and the $N_i$s are the number of degrees of freedom of each type, namely $N_0=4$, $N_{1/2}=45$, and $N_1=12$ in the SM.

Each of these can be separately computed from the Gondolo-Gelmini formula for the thermally averaged cross section \cite{Gondolo:1990dk}
\beq
\langle\sigma v\rangle=\frac{1}{8m_X^4TK_2(m_X/T)^2}\int_{4m_X^2}^\infty ds \sqrt{s}(s-4m_X^2)\sigma(s)K_1\left(\frac{\sqrt{s}}{T}\right)\label{GG}\,,
\eeq
where $\sigma(s)$ is obtained by integrating over the Mandelstam variable $t$,
\beq
\sigma(s)=\frac{-1}{16\pi s(s-4m_X^2)}\int_{t_+}^{t_-}dt|\mathcal{M}|^2\,,
\eeq
with $t_{\pm}=-(\sqrt{s/4-m_X^2}\mp\sqrt{s/4})^2$.  Specific matrix elements depend on the PIDM's spin.  To compute the cross-sections we first need the annihilation amplitudes for dark matter production from standard model particles. Using Equation (\ref{MatrixElement}), we can write down the general formula for the amplitude $\mathcal{M}$  in terms of the stress-energy tensors of dark matter and standard model particles:
 \beq\label{amplitude}
 {\cal M} = \frac{-i 8\pi G}{s} (T_{SM}^{\mu\nu}T^{DM}_{\mu\nu}-\frac{1}{2}T_{SM}T_{DM})\,,
\eeq
where $T_{SM}$ and $T_{DM}$ are the traces of the SM and DM stress-energy tensors respectively and $G \equiv m_p^{-2}$. This formula is completely general, in that it can describe the production of dark matter particles of any spin\footnote{An exception is spin 2, for which an additional, model-dependent contribution can be present.  We delay discussion of this case until the concrete model of orbifolded extra dimensions is discussed in section \ref{OKK}.} from standard model particles of any spin via graviton exchange. To obtain the particular amplitude we are interested in, we simply substitute in (\ref{amplitude}) the appropriate stress-energy tensors in momentum-space. Once we have the amplitude, we square it and sum over spin states since we are only interested in the unpolarized cross sections. 

\subsection{Scalar PIDM}

We will focus now on a scalar PIDM. For real scalar SM particles going into real scalar DM particles the relevant stress-energy tensors are:
 \begin{eqnarray}\label{scalarscalar}
T^{SM}_{\mu\nu}= \frac{1}{2}( p_{1\mu}p_{2\nu}+p_{1\nu}p_{2\mu}-\eta_{\mu\nu} p_1\cdot p_2)\nonumber\\
T^{DM}_{\mu\nu}=\frac{1}{2}( k_{a\mu}k_{b\nu}+k_{a\nu}k_{b\mu}-\eta_{\mu\nu} (k_a\cdot k_b+m_X^2))\,,
\end{eqnarray}
where the masses of the SM particles are negligible with respect to the DM mass, and so they can be treated as massless. The factor of 1/2 is necessary because we are considering real scalars, as opposed to complex scalars. The amplitude is $ {\cal M}_ {0 \rightarrow 0 } = (-i 16 \pi G/s) [(p_1 \cdot k_a)^2 + (p_1 \cdot k_b)^2-(p_1 \cdot p_2)^2]$ or, rewritten as a function of the Mandelstam variables $s=(p_1+p_2)^2$ and $t=(p_1-k_a)^2$ (and squared):
\beq
|\mathcal{M}_{0 \rightarrow 0}|^2=4G^2\pi^2\frac{(m_X^2-t)^2(m_X^2-s-t)^2}{s^2}\,.
\eeq
 For fermion SM particles going into scalar DM particles the stress-energy tensor $T^{DM}_{\mu\nu}$ remains the same, while
 \begin{eqnarray}\label{fermionscalar}
T^{SM}_{\mu\nu}= \bar{u}(p_2)[\frac{1}{4} \gamma_\mu (p_1-p_2)_\nu+\frac{1}{4} \gamma_\nu (p_1-p_2)_\mu]u(p_1)\,, 
\end{eqnarray}
and the amplitude is $ {\cal M}_ {1/2 \rightarrow 0 }(s,s') = (-i 4 \pi G/s)  \bar{u}^s(p_2)[\slashed{k_a}(p_1\cdot k_b-p_2 \cdot k_b) + \slashed{k_b}(p_1 \cdot k_a - p_2 \cdot k_a) - p_1 \cdot p_2 (\slashed{p_1}-\slashed{p2})]u^{s'} (p_1)$, where s and $s'$ label the spin states of the two incoming SM fermions. We sum over the spins s and $s'$, $|{\cal M}_{1/2 \rightarrow 0}|^2=\sum_{s,s'} |{\cal M}_{1/2 \rightarrow 0}(s,s')|^2$. 
The evaluation of the amplitude squared then reduces to the evaluation of a single trace,  $|{\cal M}_{1/2 \rightarrow 0}|^2=(16 \pi^2 G^2/s^2) \text{Tr}[\,\slashed{p_1}\,Q\,\slashed{p_2}\,Q\,]$, where we have defined the quantity  $Q=(\slashed{k_a}(p_1\cdot k_b-p_2 \cdot k_b) + \slashed{k_b}(p_1 \cdot k_a - p_2 \cdot k_a) - p_1 \cdot p_2 (\slashed{p_1}-\slashed{p_2}))$. 
We find,
\beq
|\mathcal{M}_{1/2 \rightarrow 0}|^2=-8G^2\pi^2\frac{(2m_X^2-s-2t)^2(m_X^4-2m_X^2t+t(s+t))}{s^2}\,.
\eeq
 For vector SM particles the stress-energy tensor is
 \begin{eqnarray}\label{vectorscalar}
T^{SM}_{\mu\nu}=\frac{1}{2}[\epsilon_2 \cdot \epsilon_1 (p_{1\mu} p_{2\nu} + p_{1\nu} p_{2\mu}) - \epsilon_2 \cdot p_1 (p_{2\mu} \epsilon_{1\nu} + \epsilon_{1\mu} p_{2\nu}) - \epsilon_1 \cdot p_2 (p_{1\nu} \epsilon_{2\mu} +p_{1\mu} \epsilon_{2\nu})  \nonumber\\
+(p_1 \cdot p_2+m^2_{SM})(\epsilon_{1\mu} \epsilon_{2\nu}+\epsilon_{1\nu} \epsilon_{2\mu})+\eta_{\mu\nu} (\epsilon_2 \cdot p_1 \epsilon_1 \cdot p_2 - (p_1 \cdot p_2+m_{SM}^2) \epsilon_2 \cdot \epsilon_1)] ,
\end{eqnarray}
where $\epsilon_1$ and $\epsilon_2$ are respectively the polarization vectors of the first and second SM particle, and $m_{SM}$ is their common mass. 
In the case of an exactly massless photon-like vector particle, the trace part $-(1/2) T_{SM}T_{DM}$ is zero and the amplitude is
 \begin{eqnarray}\label{Mvectorscalar}
\mathcal{M}^{m_{SM}=0}_{1 \rightarrow 0}=\frac{1}{2}[ \epsilon_2 \cdot \epsilon_1 (2 p_1 \cdot k_a p_2 \cdot k_b+2p_1 \cdot k_b p_2 \cdot k_a - 2 (p_1p_2)^2 ) \nonumber\\
 - \epsilon_2 \cdot p_1 (2 p_2 \cdot k_a \epsilon_1 \cdot k_b + 2p_2 \cdot k_b \epsilon_1 \cdot k_a - 2 p_2\cdot \epsilon_1p_1\cdot p_2) \nonumber\\
 -  \epsilon_1 \cdot p_2 (2 p_1 \cdot k_b \epsilon_2 \cdot k_a + 2p_1 \cdot k_a \epsilon_2 \cdot k_b - 2 p_1\cdot \epsilon_2 p_1\cdot p_2)  \nonumber\\
+ p_1\cdot p_2(2\epsilon_1 \cdot k_a \epsilon_2\cdot k_b+2\epsilon_1 \cdot k_b \epsilon_2\cdot k_a - 2\epsilon_1 \cdot \epsilon_2 p_1 \cdot p_2)  \nonumber\\
+ (-2k_a \cdot k_b - 4m_X^2)(\epsilon_2\cdot p_1 \epsilon_1 \cdot p_2 - p_1\cdot p_2 \epsilon_2 \cdot \epsilon_1)].
\end{eqnarray}
In analogy with the spin 1/2 case, we have to square the amplitude and sum over polarization states using $\sum \epsilon_{1\mu}^* \epsilon_{1\nu}=\sum \epsilon_{2\mu}^* \epsilon_{2\nu}=-g_{\mu\nu}$. 
We arrive at the final expression for the amplitude squared
 \beq
|\mathcal{M}^{m_{SM}=0}_{1 \rightarrow 0}|^2=8G^2\pi^2\frac{(m_X^4-2m_X^2t+t(s+t))^2}{s^2}.
\eeq
If the SM vectors are massive, the trace part is no longer zero and we have to add the corresponding terms to (\ref{Mvectorscalar}), which are proportional to $m_{SM}^2$. The spin sum identities are modified to $\sum \epsilon_{1\mu}^*(p_1) \epsilon_{1\nu}(p_1)=-g_{\mu\nu}+p_{1\mu}p_{1\nu}/m_{SM}^2$ and $\sum \epsilon_{2\mu}^*(p_2) \epsilon_{2\nu}(p_2)=-g_{\mu\nu}+p_{2\mu}p_{2\nu}/m_{SM}^2$ for a massive particle.
As before, we square the expression for the amplitude, which now contains terms proportional to the mass of the SM particles, and use the sum identities above. Only at the end of the calculation we take the limit $m_{SM}\to 0$. We find the simple result
 \beq
|\mathcal{M}^{m_{SM}\neq0}_{1 \rightarrow 0}|^2=|\mathcal{M}^{m_{SM}=0}_{1 \rightarrow 0}|^2+|\mathcal{M}_{0 \rightarrow 0}|^2\,.
\eeq

This is consistent with the Goldstone boson equivalence theorem. Indeed, suppose that we are above the electroweak symmetry breaking scale. Then the $SU(3)\times SU(2)\times U(1)$ symmetry group is unbroken, the Higgs field is in its symmetric ground state and the vector degrees of freedom of the standard model are exactly massless. The Higgs boson contributes 4 scalar d.o.f. to the annihilation amplitude $|\mathcal{M}_{0 \rightarrow 0}|^2$ and the vector bosons contribute 4 massless transverse d.o.f. to the annihilation amplitude $|\mathcal{M}^{m_{SM}=0}_{1 \rightarrow 0}|^2$. Instead, if we are below the electroweak scale, the symmetry is spontaneously broken and 3 of the 4 scalar d.o.f. are eaten by the W and Z vector bosons, which become massive. Then the Higgs field contributes 1 scalar d.o.f. to $|\mathcal{M}_{0 \rightarrow 0}|^2$, while the photon contributes 1 massless transverse d.o.f. to $|\mathcal{M}^{m_{SM}=0}_{1 \rightarrow 0}|^2$ and the 3 massive vectors contribute to $|\mathcal{M}^{m_{SM}\neq0}_{1 \rightarrow 0}|^2$ with one longitudinal and two transverse d.o.f. Since $|\mathcal{M}^{m_{SM}\neq0}_{1 \rightarrow 0}|^2$ is just the sum of the massless vector amplitude and the scalar amplitude in the high energy limit, the end result is the same, as it should be. 

Given the amplitudes for scalar PIDM production,  equation (\ref{GG}) can be used to arrive at the thermally averaged cross sections:
\begin{eqnarray}\label{sigv}
\langle\sigma v\rangle_0&=&\frac{\pi m_X^2}{8 m_p^4}\left[\frac35\frac{K_1^2}{K_2^2}+\frac25+\frac45\frac{T}{m_X}\frac{K_1}{K_2}+\frac{8}{5}\frac{T^2}{m_X^2}\right]\,,\nonumber\\
\langle\sigma v\rangle_{1/2}&=&\langle\sigma v\rangle_{1} \ = \  \frac{4 \pi T^2}{m_p^4}\left[\frac{2}{15}\left(\frac{m_X^2}{T^2}\left(\frac{K_1^2}{K_2^2}-1\right)\right.\right. \nonumber\\
&& {} \left.\left. + 3\frac{m_X}{T}\frac{K_1}{K_2}+6\right)\right]\,.\nonumber
\end{eqnarray}
  The $K_i$ are modified Bessel functions, with argument $m_X/T(a)$, and the brackets asymptote to 1 for $m_X\gg T$, leaving the prefactor to display the non-relativistic behavior ($s$-wave for scalars, and $d$-wave otherwise).  

\begin{figure}[h]
\begin{centering}
\includegraphics[width=12cm]{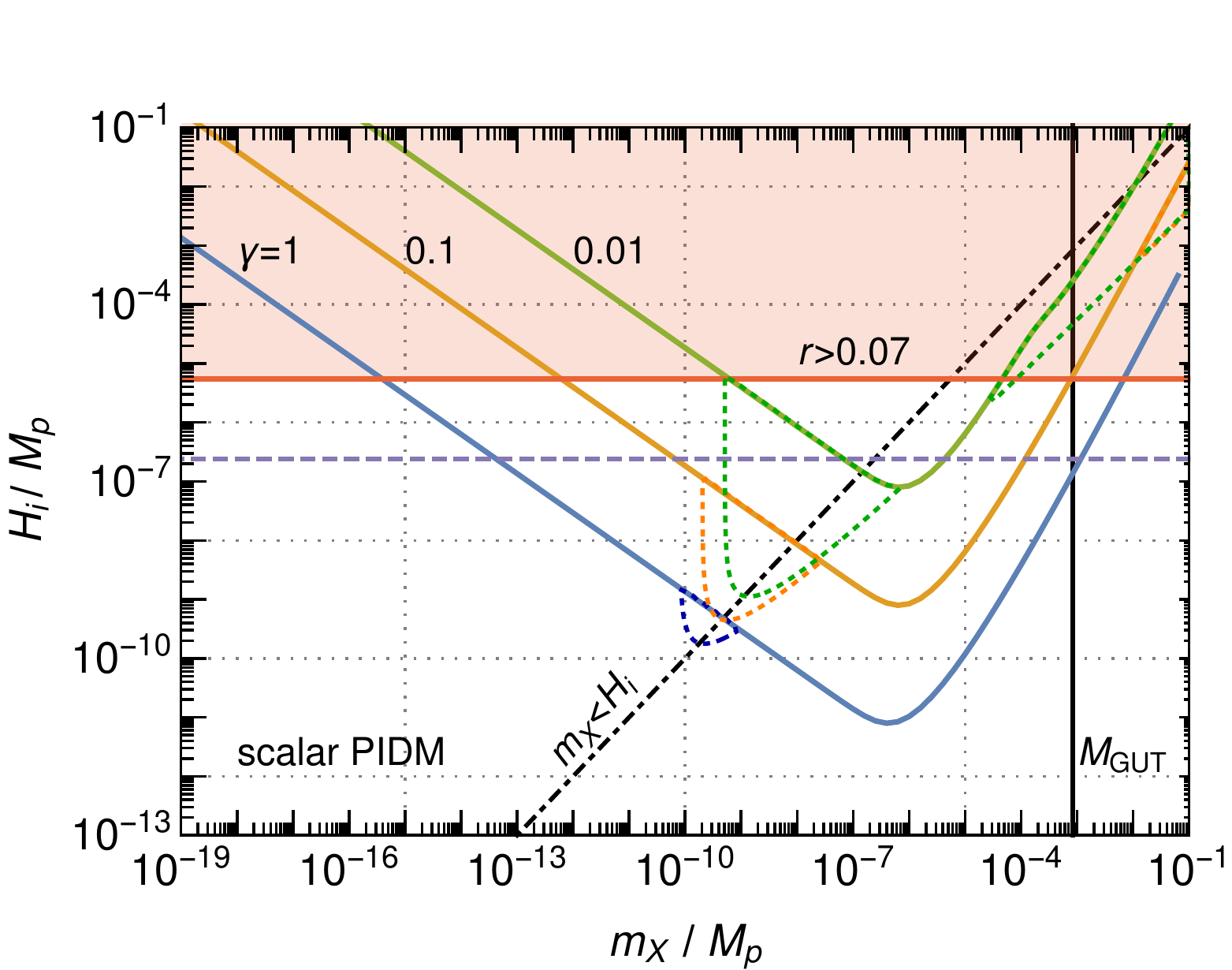}
\caption{The value of the Hubble rate at the start of reheating that gives the observed dark matter abundance, as a function of the mass of the PIDM, for various values of reheating efficiency (blue: $\gamma=1$, orange: $\gamma=0.1$, and green: $\gamma=0.01$).  The red region is excluded from the current bound on the tensor-to-scalar ratio, and the purple dashed line is the projected sensitivity for future CMB experiments, from \cite{Errard:2015cxa}. The dotted lines show the additional `gravitational production' contribution \cite{Chung:2001cb,Kuzmin:1999zk, Chung:2004nh}. The dashed-dotted line marks $m_X=H_i$, left of which is excluded for scalar PIDMs (unless corrections to the PIDM potential are important during inflation).  All values are given in units of $m_p$.}
\label{hofm}
\end{centering}
\end{figure}

The thermally averaged cross sections  can be used in equation (\ref{Xfint}) to find the value of the relic abundance numerically. If the PIDM is to be the dominant component of the dark matter, we require the parameters to be such that the abundance measured by Planck, $\Omega_c h^2=0.1198 \pm 0.0015$ (68\%CL) \cite{Ade:2015xua} is produced, which will define a lower dimensional manifold in parameter space for which this occurs.  One useful way to express this is illustrated in Fig.\,\ref{hofm} whose level curves are the lines in the $H_i-m_X$ plane which satisfy this criterion, for several values of the reheating efficiency $\gamma$. For very efficient reheating, $\gamma\approx1$, the PIDM is produced shortly after reheating via freeze-in. Only for smaller values of $\gamma$, does the contribution to production during reheating become relevant. 

The shape of the level curves interpolate between the analytic estimates of section \ref{AbC} for small and large masses.  The Hubble rate must be larger to produce the same cross section for small masses due to the suppression of the cross section in this regime, and similarly for large masses due to the exponential Boltzmann suppression. We also include the contribution from `gravitational production' discussed in section \ref{otherprod} as the dotted lines, which is relevant for $m_X \simeq H_i$.  The results are robust when varying $w$ in the range $(-2/3,2/3)$. 

We see that, in the limit of instantaneous reheating ($\gamma=1$), the range of viable PIDM masses is quite large, spanning from $10^{-10}-10^{-2} m_p$.  However, this range is very sensitive to the reheating efficiency, and shrinks significantly if reheating takes just a bit longer. If $\gamma\lesssim 10^{-3}$, which corresponds to a reheating lasting $N_{rh}\gtrsim 10/(1+w)$ $e$-folds, the freeze-in mechanism cannot be operational for any value of the PIDM mass.  For values just above this, the allowed mass range is centered at $m_X\sim 10^{-6}m_p$.

However, the philosophy of our scenario prefers substantially higher PIDM masses, as these are more minimal, especially in the scalar case where quantum corrections typically will drive the mass towards the ultraviolet cutoff in the effective theory.  Indeed, for large enough $\gamma$, $m_X\gtrsim M_{GUT}$ is in accordance with observations. The maximum value of the PIDM mass for a given bound on the tensor to scalar ratio is
\beq
m_{\text{max}}=0.023\,\gamma^{1/2}r^{1/4}m_p\,.
\eeq

If we take the current bounds, $r<0.07$, then the maximum allowed mass is $0.013 m_p$.  This value decreases if the reheating efficiency is smaller.  If we favor large masses, therefore, the scale of inflation must be such that we expect to see tensor modes in the next round of CMB experiments.  For the futuristic sensitivity of $r\sim 10^{-4}$ quoted in \cite{Errard:2015cxa}, the maximum mass can be improved by almost an order of magnitude. Remarkably, the entire region of parameters with GUT scale masses can be probed in the foreseeable future.

 Therefore, if the next generation of CMB experiments exclude primordial tensor modes to this level, the PIDM scenario will only be viable if its mass is significantly below the natural cutoff scale.  Let us stress again that our conclusions are predicated on the standard reheating setup, with a constant equation of state and decay rate.  More general scenarios may alter these conclusions, but at the cost of introducing additional model dependence.

\subsection{Fermion PIDM}
\begin{figure}[h]
\begin{centering}
\includegraphics[width=12cm]{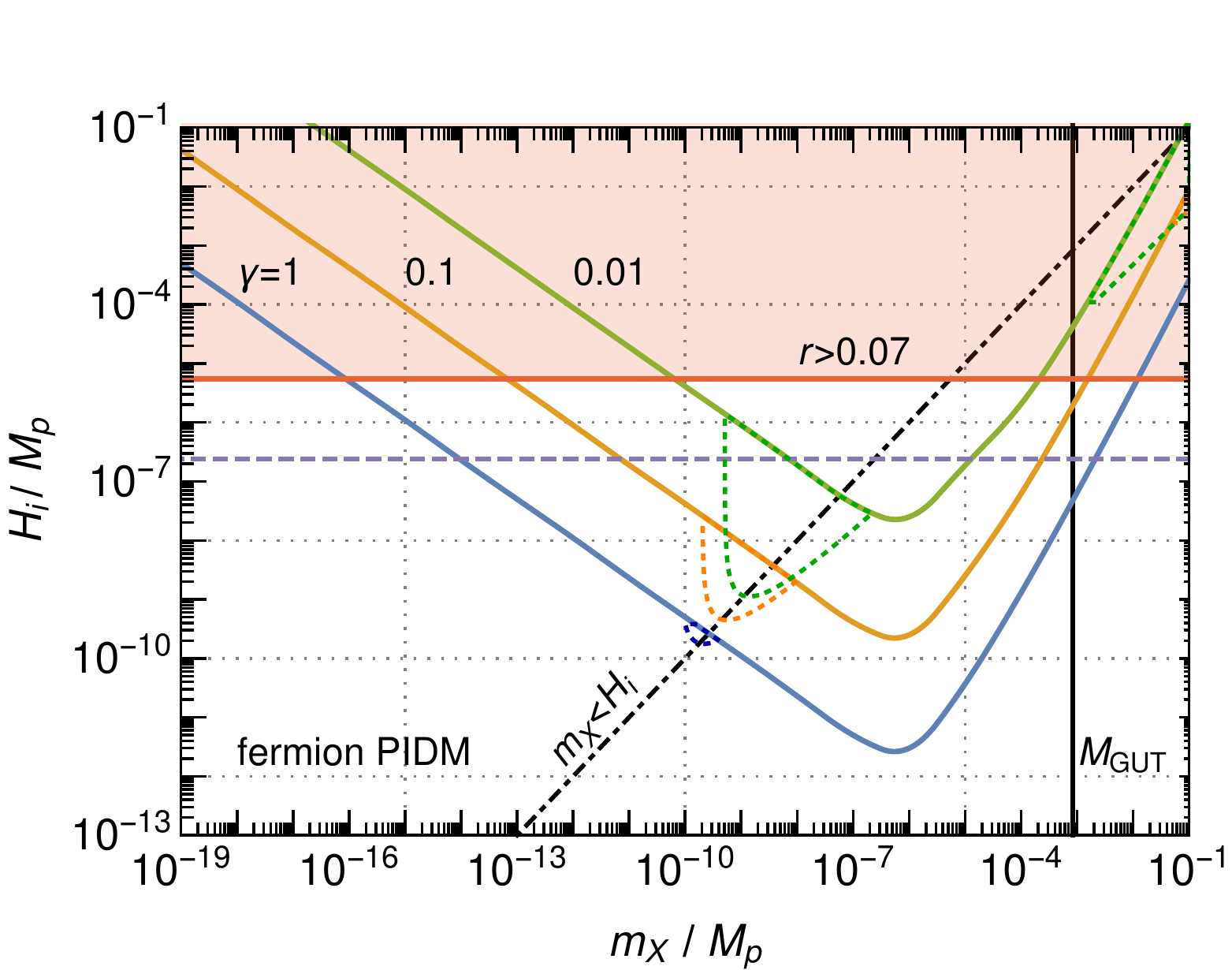}
\caption{Same as Fig. \ref{hofm}, but for fermion PIDMs.  The blue curve is for $\gamma=1$, orange for $\gamma=0.1$ and green for $\gamma=0.01$. The region above the red line is excluded from the current bound on the tensor-to-scalar ratio. 
Note that, in contrast to scalar PIDM, also values $m_X<H_i$ (gray line) are allowed.}
\label{fermionplot}
\end{centering}
\end{figure}

Very little changes when going from a scalar PIDM to a fermion PIDM. The only difference is that the DM stress-energy tensor is now $T^{DM}_{\mu\nu}= \bar{u}(k_b,m_X)[\frac{1}{4} \gamma_\mu (k_a-k_b)_\nu+\frac{1}{4} \gamma_\nu (k_a-k_b)_\mu]u(k_a,m_X)$.  For SM scalars going into fermion DM equation  (\ref{amplitude}) yields
\beq
|\mathcal{M}_{0 \rightarrow 1/2}|^2=8G^2\pi^2\frac{(m_X^2-t)(-m_X^2+s+t)(-2m_X^2+s+2t)^2}{s^2},
\eeq
while for fermion SM particles
\bea
|\mathcal{M}_{1/2 \rightarrow 1/2}|^2=\frac{8G^2\pi^2}{s^2} [32m_X^8-32m_X^6(s+4t) +2m_X^4(5s^2+64st+96t^2) \nonumber\\
-4m_X^2(s^3+13s^2t+40st^2+32t^3)+s^4+10s^3t+42s^2t^2+64st^3+32t^4],
\eea
and finally for vector SM bosons
\beq
|\mathcal{M}_{1 \rightarrow 1/2}|^2=-32G^2\pi^2\frac{(m_X^4-2m_X^2t+t(s+t))[s^2+2(m_X^4-2m_X^2t+t(s+t))]}{s^2}.
\eeq
Using (\ref{GG}) we obtain the thermally averaged cross sections:
\begin{eqnarray}
\langle\sigma v\rangle_0&=&\frac{ \pi m_X T}{2 m_p^4}\left[\frac{4}{5}\frac{T}{m_X} +\frac{1}{5}\frac{m_X}{T}-\frac{1}{5}\frac{m_X}{T}\frac{K_1^2}{K_2^2}+\frac{2}{5}\frac{K_1}{K_2}\right]\,,\nonumber\\
\langle\sigma v\rangle_{1/2}=\langle\sigma v\rangle_{1}&=&\frac{4 \pi m_X T}{m_p^4}\left[\frac{6}{5}\frac{T}{m_X} +\frac{2}{15}\frac{m_X}{T}-\frac{2}{15}\frac{m_X}{T}\frac{K_1^2}{K_2^2}+\frac{3}{5}\frac{K_1}{K_2}\right]\,.\nonumber\\
\end{eqnarray}
As before, the argument of the Bessel functions is $m_X/T$, and the brackets asymptote to 1 for  $m_X\gg T$.  In this scenario, all the cross sections are $p$-wave. The level curves in Fig.\,\ref{fermionplot} show the lines for which the fermion PIDM abundance matches the cold dark matter density measured by Planck.
As one can see, the results are very similar to the case of a scalar PIDM. In particular, in the limit of instantaneous reheating the allowed range for the DM mass is quite large, with $0.01 m_p$ being the largest possible mass compatible with the constraint on the inflation scale. If we lower $\gamma$ the viable mass range quickly shrinks. The upper bound on $m_X$ is $10^{-3}m_p$ for $\gamma=0.1$ and $10^{-4}m_p$ for $\gamma=0.01$. The main point remains that the minimal PIDM scenario demands a high inflation scale and fast reheating.

\subsection{Vector PIDM}
\begin{figure}[h]
\begin{centering}
\includegraphics[width=12cm]{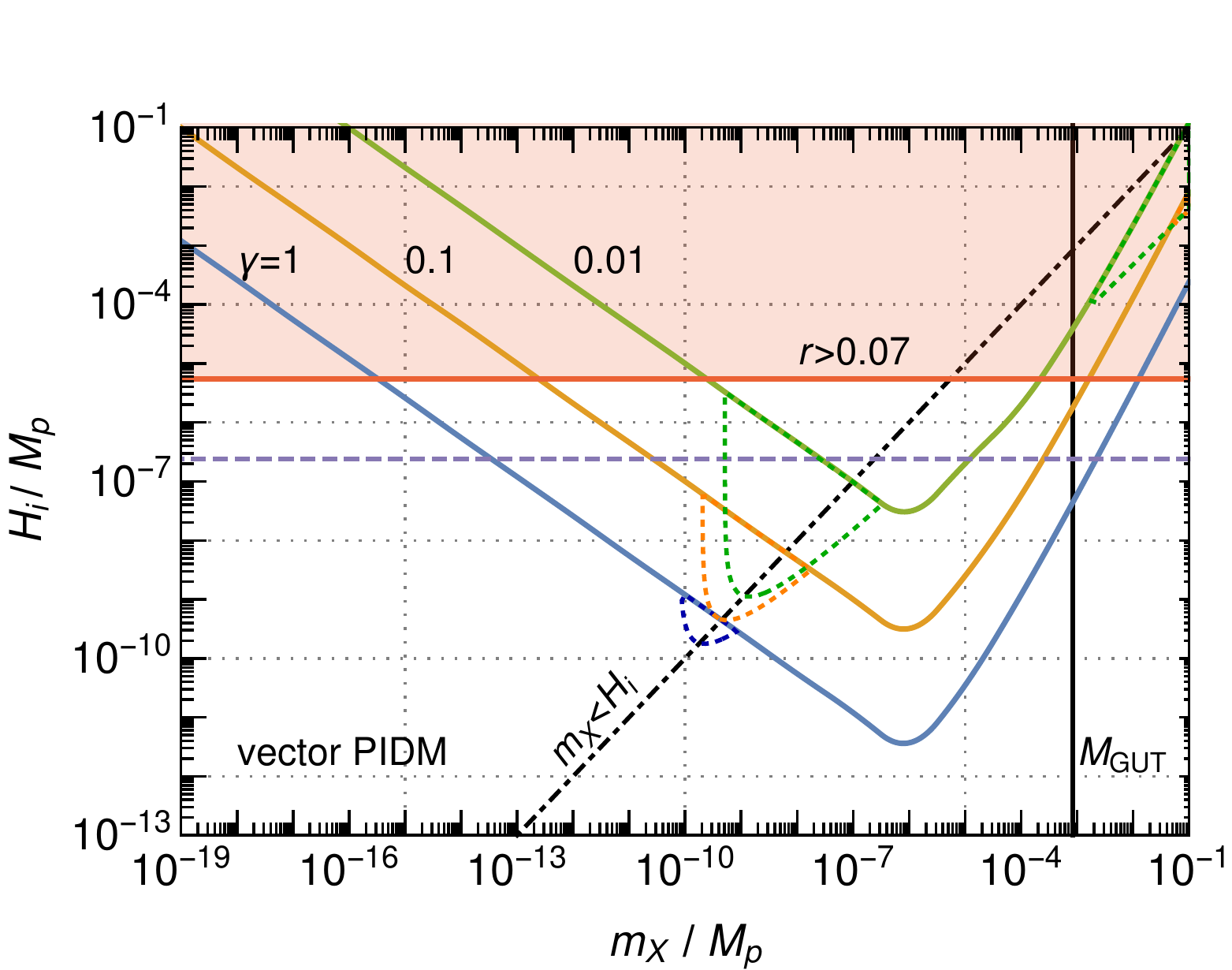}
\caption{Same as Fig. \ref{hofm}, but for vector PIDMs.    The blue curve is for $\gamma=1$, orange for $\gamma=0.1$ and green for $\gamma=0.01$. The region above the red line is excluded from the current bound on the tensor-to-scalar ratio. 
}
\label{vectorplot}
\end{centering}
\end{figure}

In the case where the PIDM is a vector, the squared amplitude for scalar production is 
\bea
|\mathcal{M}_{0 \rightarrow 1}|^2=\frac{12 \pi ^2 G^2 \left(m_X^2-t\right)^2 \left(-m_X^2+s+t\right)^2}{s^2} = 3 |\mathcal{M}_{0 \rightarrow 0}|^2.
\eea
For production by SM fermions, it is
\bea
|\mathcal{M}_{1/2 \rightarrow 1}|^2=-\frac{2 \pi ^2 G^2}{s^2} [12 m_X^8-12 m_X^6 (s+4 t)+m_X^4 \left(5 s^2+48 s t+72 t^2\right)-2m_X^2 \nonumber\\
 \left(2 s^3+11 s^2 t+30 s t^2+24 t^3\right)+t \left(5 s^3+17 s^2 t+24 s t^2+12t^3\right)],
\eea
and for production by SM vectors, it is
\beq
|\mathcal{M}_{1 \rightarrow 1}|^2=\frac{2 \pi ^2 G^2 \left(m_X^4-2 m_X^2 t+s^2+s t+t^2\right) \left(3 m_X^4-6 m_X^2 t+s^2+3 s t+3 t^2\right)}{s^2}.
\eeq
The thermally averaged cross sections are:
\begin{eqnarray}
\langle\sigma v\rangle_0&=&\frac{3 \pi m_X^2}{8 m_p^4}\left[\frac{3}{5}\frac{K_1^2}{K_2^2}+\frac{2}{5}+\frac45\frac{T}{m_X}\frac{K_1}{K_2}+\frac{8}{5}\frac{T^2}{m_X^2}\right]\,,\nonumber\\
\langle\sigma v\rangle_{1/2}&=&\langle\sigma v\rangle_{1} \ = \frac{\pi m_X^2}{m_p^4}\left[\frac{11}{20}\frac{K_1^2}{K_2^2}+\frac{9}{20}+\frac{13}{20}\frac{T}{m_X}\frac{K_1}{K_2}+\frac{13}{10}\frac{T^2}{m_X^2}\right],
\end{eqnarray}
with the argument of the Bessel functions $m_X/T$, and brackets which asymptote to 1 for $m_X\gg T$.  The cross sections have the same structure as the scalar-to-scalar one, with different prefactors. In particular, $\langle\sigma v\rangle_0$ is just 3 times the scalar-to-scalar cross section, due to the 3 polarization states of a massive vector particle. Fig.\,\ref{vectorplot} shows the Hubble parameter at the end of inflation that gives the correct relic abundance as a function of the PIDM mass. 

\subsection{Nonminimally Coupled PIDM}\label{nmPIDM}
In the case where the PIDM is a scalar, we may be interested in the scenario where the theory contains an additional nonminimal coupling to gravity, of the form
\beq\label{NMC}
 \mathcal{L}_{\text{NM}}=\frac12 (\xi_\phi \phi^2  +  \xi_X X^2 ) R
\eeq
where $X$ is the PIDM, and $\phi$ is any scalar present in the standard model, each with their own coupling.  The matrix element for this process is 
\beq
|\mathcal{M}|^2=4 G^2 \pi^2 \left(\frac{m^4 + s t + t^2 - m^2 (s + 2 t + 2 s \xi_\phi) -
   s^2 (\xi_X + \xi_\phi + 6 \xi_X \xi_\phi)}{s}\right)^2
\eeq
leading to the thermally-averaged cross section
\bea
\langle\sigma v\rangle=\frac{G^2 m^2 \pi}{360 x^2} \bigg(x^2 (2 +
      5 (1 + 2 X_c + 2 X_c^2) X_h^2) \frac{K_1^2(x)}{K_2^2(x)} +
   6 x (1 + 5 X_c^2 X_h^2) \frac{K_1(x) }{K_2(x)} +\nonumber\\
   2 (6 + 30 X_c^2 X_h^2 + x^2 (-1 + 5 X_c X_h^2 + 5 X_c^2 X_h^2))\bigg)
\eea
where we have defined $X_h=1+6 \xi_\phi$, $X_c=1+6 \xi_X$, and $x=m/T$.  The nonrelativistic limit of this expression is
\beq
\langle\sigma v\rangle\rightarrow \frac{G^2 m^2 \pi}{8} (1 + 4 \xi_X)^2 (1 + 6 \xi_\phi)^2\,.
\eeq
Therefore, the nonminimal coupling can be effectively taken into account by the replacement $m\rightarrow m(1 + 4 \xi_X)(1 + 6 \xi_\phi)$ for the scalar-scalar process.  If the standard model Higgs has a large nonminimal coupling to gravity, then it will dominate the production cross section for the PIDM.  If either value of the nonminimal coupling is significantly larger than ${{\cal O}(1)}$, it becomes harder for the minimal PIDM scenario, in which the mass is GUT scale, to be realized, as the effective mass can become several orders of magnitude larger than the actual mass. On the other hand, our conclusions so far are robust when allowing for a small non-minimal coupling,
in agreement with the expected size generated by renormalization group effects \cite{Herranen:2014cua}.

\section{In Situ PIDMs}\label{insitu}

In the previous section we have demonstrated that the PIDM is a viable dark matter candidate for large reheating temperatures, and that it is rather insensitive to the details of the scenario, other than the mass of the PIDM.  In this section, we outline several scenarios that naturally contain particles capable of acting as PIDMs.

\subsection{Orbifold KK PIDMs}\label{OKK}
\begin{figure}[h]
\begin{centering}
\includegraphics[width=12cm]{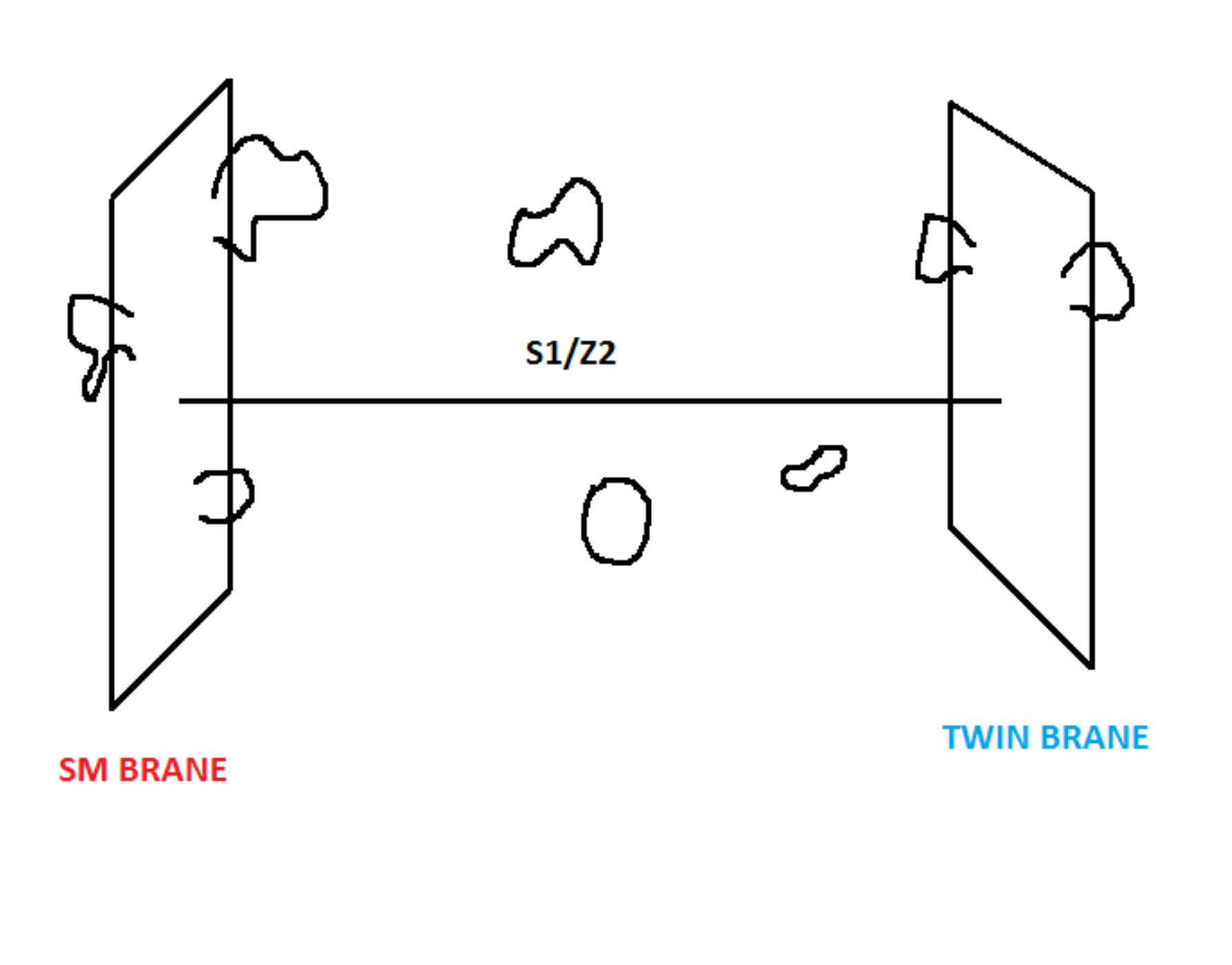}
\caption{Minimal setup in which the PIDM is the first Kaluza-Klein excitation of the graviton in a $S_1/Z_2$ orbifold compactification. The orbifold action $Z_2$ explicitly breaks KK number conservation, but leaves an invariance under KK parity, which makes the lightest KK graviton stable. The SM fields (open strings in the figure) live on a brane stuck at one of the two orbifold fixed points. The $Z_2$ symmetry then requires the presence of an identical brane located at the opposite fixed point. The graviton (the closed strings in the figure) is the only field able to propagate in the extra dimension.}
\label{setup}
\end{centering}
\end{figure}

The possibility that a Kaluza-Klein (KK) resonance may act as the dark matter particle is an attractive scenario \cite{Cheng:2002ej}.  Here, the 4d mass is a consequence of the momentum in the extra dimension(s), which is quantized due to their compact nature. Since the 4d mass is inversely proportional to the compactification size, the implementation of a minimal PIDM (GUT scale mass) in this scenario requires that the extra dimensions are small, e.g. related to the GUT scale.  We consider the simplest set-up depicted in Fig.\ref{setup}, starting from a 5d Minkowski space where one of the spatial dimensions is compactified on the orbifold segment $S_1/Z_2$. While the orbifold breaks KK number conservations, the lightest KK state is stable due to the KK parity symmetry  \cite{Feng:2003nr}. The 4d and 5d Planck masses are related by
\beq
M_5^3=m_p^2/R
\eeq
where $M_5$ and $m_p$ are the 5d and 4d Planck scales, respectively, and $R$ is the compactification scale (the length of the segment). 

Upon compactification the 5d metric $g_{MN}$ with $M,N=0,..,4$ is split into three components: a collection of 4d tensor fields $g^{\left(n\right)}_{\mu\nu}$, a collection of 4d vector fields $g^{\left(n\right)}_{\mu4}$ and a collection of 4d scalar fields $g^{\left(n\right)}_{44}$, one for each $n\in \mathbb{N}$ and with $\mu,\nu=0,...,3$. Each of these three components contains a massless (zero) mode and an infinite KK tower of massive states, with masses that are integer multiples of the inverse compactification radius: $m_n=n/R$. So, for example, the modes with $n=0$ correspond to the massless graviton $g^{\left(0\right)}_{\mu\nu}$, a $U(1)$ gauge boson $g^{\left(0\right)}_{\mu4}$ and the radion $g^{\left(0\right)}_{44}$, while the states with $n\ne0$ are their massive counterparts. A massless graviton in five dimensions has 5 degrees of freedom, and this number matches the counting of degrees of freedom at each KK level $n$. For $n=0$ we have a massless graviton in four dimensions, a massless gauge boson and a massless scalar field for a total of 2+2+1=5 degrees of freedom. For $n\neq0$ the counting is slightly more involved. 
For a $S_1$ compactification, one can see that the infinite dimensional symmetry of $S_1$ is spontaneously broken by the vacuum configuration $g_{MN}=\eta_{MN}$, giving rise to a Higgs-like mechanism which is geometrical in nature \cite{Aulakh:1985un}. As a result, the massless spin-2 fields $g^{\left(n\right)}_{\mu\nu}$ absorb the vector and scalar fields at the same KK level $n\neq0$ and become massive. A massive graviton in 4d has 5 degrees of freedom, and the two numbers match also in this case. 

In order to make a more realistic scenario for particle physics, our setup has an additional ingredient, the orbifolding by $Z_2$. The $Z_2$ breaks KK-number conservation, but still leaves a conserved KK-parity that guarantees the stability of the lightest KK resonance, which will be our PIDM candidate. Note that KK-parity is not the same as parity under the $Z_2$, but can be expressed as a conservation of  the parity $(-1)^{KK}$, where $KK$ is the KK-number. This means that an odd KK-number state cannot decay into SM particles, which carry even KK-number/parity.

Given the circle compactification with coordinate $x^4\equiv y\in [0,2\pi R]$, the segment orbifold is obtained by identifying $y$ and $-y$, with the two fixed points at $y=0$ and $y=\pi R$. We decide to have the SM fields trapped on a brane located at one of the fixed points, with the 5d graviton the only field able to propagate in the bulk. Fields defined on the orbifold can be even or odd under $y\rightarrow-y$, depending on whether they satisfy Dirichlet or Neumann boundary conditions. An even field satisfies $\Phi_+(y)=\Phi_+(-y)$, while for an odd field $\Phi_-(y)=-\Phi_-(-y)$. Expanding the fields in a Fourier series in the compact coordinate $y$ we get
\begin{eqnarray}
\Phi_+(x,y)\sim \Phi_+^{\left(0\right)}(x)+\sum_{n=1}^\infty  \Phi_+^{\left(n\right)}(x)\cos(ny/R) \nonumber \\
\Phi_-(x,y)\sim \sum_{n=1}^\infty  \Phi_-^{\left(n\right)}(x)\sin(ny/R) \nonumber\\
\end{eqnarray}                        
with mass $m_n=n/R$ for the $n$th mode. It is clear from the expansion that only even fields contain a zero mode. Consider now the 5d metric $g_{MN}(x,y)$. Given that $\partial_{\mu}$ is even and $\partial_{4}$ is odd under the $Z_2$ parity, coordinate invariance of the gravity action actually tells us that $g_{\mu\nu}$ and $g_{\mu4}$ must have opposite parities, whereas  $g_{\mu\nu}$ and $g_{44}$ have the same parity. This means that if  $g_{\mu\nu}$ and $g_{44}$ are even (we need  $g_{\mu\nu}$ to be even, otherwise there would be no massless graviton), then $g_{\mu4}$ is odd and does not have a zero mode. In other words, we can use the $Z_2$ symmetry to project out the unwanted gauge boson. That still leaves us with an unstabilised scalar field, the radion  $g^{\left(0\right)}_{44}\equiv \phi(x)$, which controls the size of the extra dimension. There are various ways to stabilise the radion. Moduli stabilisation is not the focus of this paper, therefore we will limit ourselves to pointing out a couple of ways in which this can be achieved in our scenario. One, very minimal, possibility is to have the radion stabilised by an inter-brane potential. Another possibility, if we embed our model in a larger 10 dimensional string theory compactification, is moduli stabilisation by fluxes and non-perturbative effects in string theory. 

To sum up, the particle content of our model is the following: a massless graviton, a stabilised (massive) scalar field and an infinite KK tower of massive gravitons, all free to propagate in the bulk, plus all the standard model fields which are confined on the brane. The gravitons interact both with the two branes and among themselves.

\begin{figure}
\centering
\begin{subfigure}{.45\textwidth}  
\centering
\includegraphics[width=7cm]{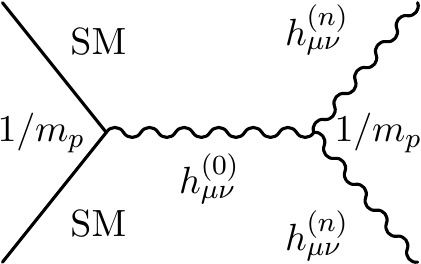}
\end{subfigure}   
\begin{subfigure}{.45\textwidth}
\centering
\includegraphics[width=4.5cm]{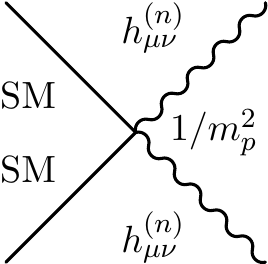}
\end{subfigure}
\caption{\label{contact}Pair production of KK gravitons via graviton exchange (left) and contact interaction (right) diagrams. This process is possible
both for even and odd $n$, and the amplitude scales as $1/m_p^2$. For $n=1$ this process describes one of two most relevant production channels of the
lightest KK graviton.}
\end{figure}

The dimensionally reduced quadratic Lagrangian for the massive KK modes of the graviton has the Fierz-Pauli form:
\beq
\mathcal{L}^n=\frac{1}{2}\left(\partial_\lambda h^{\mu\nu,n}\partial^\lambda h_{\mu\nu}^n-\partial_\lambda h^{\mu,n}_\mu \partial^\lambda h^{\nu,n}_\nu-2\partial_\lambda h^{\lambda\nu,n}\partial^\mu h_{\mu\nu}^n+2\partial^\nu h^{\lambda,n}_\lambda \partial^\mu h_{\mu\nu}^n+m_n^2(h^{\mu,n}_\mu h^{\nu,n}_\nu - h^{\mu\nu,n}h_{\mu\nu}^n) \right),
\eeq  
where $h_{\mu\nu}$ is the linearized 4d metric. The higher order terms describe the self-interactions of the KK gravitons. The interaction term with the SM fields is
\beq
\mathcal{L}_{int}^n=\frac{\sqrt{8\pi}}{2m_p}h_{\mu\nu}^nT^{\mu\nu}_{SM}.
\eeq
Now we are ready to compute the abundance of the lightest KK excitation of the graviton in our setup. For simplicity, we will call the nth KK mode of the graviton $KK_n$, the stable  $KK_1$ being the PIDM candidate. 
First of all, since the dark matter particle in this setup has odd KK parity, it cannot directly couple to the standard model sector, whose particles have even KK parity. This means that it can only be produced in pairs. Double production of KK modes takes place through a graviton exchange diagram as well as a contact interaction diagram, shown in Fig.~\ref{contact}, which we add up to obtain the full amplitude. In principle we should consider the contribution coming from all the diagrams of this type with a generic combination of higher order same-parity KK modes in the final state, which can then decay to the stable $KK_1$ and add to its abundance.  In the limit of natural PIDM mass, however, we can safely ignore those higher order contributions and only consider the diagrams corresponding to the production of $KK_1$. Moreover, we can assume that the mediator in the first diagram is just the massless graviton, as the other terms coming from diagrams in which the mediator is a massive graviton will be subleading in this regime. 

To compute the amplitudes we need, apart from the usual SM-graviton vertex that we already used in the other scenarios, the triple graviton vertex and the two graviton seagull vertex that gives rise to the contact interaction. We use the Feynman rules of \cite{Bjerrum-Bohr:2014lea} for the triple graviton vertex $\tau^{\mu\nu}_{\alpha\beta,\gamma\delta}(k_a,k_b)$ and the contact interaction vertex $V^{\mu\nu,\alpha\beta}_S(p_1,p_2)$, with $S=0,1/2,1$ depending on the spin of the SM particles,  $p_1$,$p_2$ the momenta of the incoming SM particles and $k_a$,$k_b$ the momenta of the outgoing gravitons. Strictly speaking, the Feynman rules for the two vertices are only valid for massless gravitons, but we can get the Feynman rules for massive gravitons by dimensional reduction of the corresponding 5d vertices. We know that the tensor structure of the vertices is the same in any number of dimensions, therefore we can use the same rules but with 5d indices and momenta. The rules are now exact because we have pure Einstein gravity with massless gravitons in the extra-dimensional space. To go from 5d to 4d we just contract the 5d vertices with the 4d massive graviton polarization tensors in order to project out the unwanted degrees of freedom, and we express the 5-momenta in terms of the 4d ones. So for example the product $k_a^{(5)} \cdot k_b^{(5)}$ of the massless gravitons 5-momenta becomes after dimensional reduction: $k_a^{(5)} \cdot k_b^{(5)}=k_a^{(4)} \cdot k_b^{(4)}+m_am_b$, where $m_a$ and $m_b$ are the masses of the two KK gravitons in the final state, which correspond to the components of the 5-momenta in the extra dimension (see Appendix \ref{app:3g} for details).

The analytical expressions for the amplitudes are rather complicated and uninteresting, so we will not show them here. We will confine our discussion here to their scaling behavior, since this contains the relevant physics. The amplitudes scale as $|\mathcal{M}|^2\sim E^{12}/(m^8m_p^4)$, where $E$ represents the typical energy scale of the process and $m$ is the graviton mass.  A peculiar property of these amplitudes, then, is that they diverge for $m\rightarrow 0$, signaling that the theory becomes strongly coupled at low masses. This is a well-known feature of the theory of a single massive graviton \cite{ArkaniHamed:2002sp}. In fact, for a single graviton in four dimensions with a Planck scale $m_p$ and a mass $m$, one can show that there is a sensible effective field theory which is valid up to a low-energy cutoff $\Lambda=(m^2m_p)^{1/3}$ (low with respect to the Planck scale), above which the EFT becomes invalid. Looking at the scaling behaviour of the amplitudes, unitarity is lost, and the EFT breaks down, when the ratio $E^{12}/(m^8m_p^4)$ is of order one. This condition sets an energy scale in our theory, which is exactly the cutoff $\Lambda$. 
It is useful to look at the problem from the point of view of 5d Einstein gravity, which we know is a consistent, stable and cutoff free theory. After compactification, 5d gravity is equivalent to the 4d theory of an infinite number of fully interacting massive gravitons. It is clear from this perspective that there should be no strong coupling problem or low scale cutoff in the complete 4d theory, which should be valid all the way up to the 5d Planck mass. What happens is that the 4d graviton modes interact in such a way as to cancel out all the strong coupling effects and the low energy cutoff disappears from the final theory \cite{Schwartz:2003vj,Hinterbichler:2011tt}. Any truncation of the full 4d theory to a finite number of KK modes will automatically introduce a cutoff in the effective field theory.

\begin{figure}
\begin{center}
\includegraphics[width=0.3\textwidth]{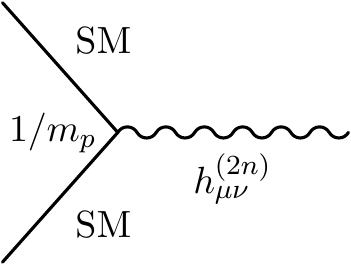}
\end{center}
\caption{\label{KK_invdecay}Production of single KK gravitons with even KK number via inverse decay of SM particles. The amplitude scales with the first power $1/m_p$
of the Planck mass.}
\end{figure}

Given that the amplitudes and cross sections are extremely cumbersome and difficult to work with, it is useful to obtain approximate expressions when the high mass approximation is valid. If the maximum temperature during reheating, Eq.\,\ref{maxtem}, is much higher than the mass of the PIDM particle, or equivalently the inverse compactification radius, then our approximation is bound to fail, since higher order KK modes will be produced copiously. Roughly, the number of excited KK modes $N$ during reheating is $N \sim T_{max}/m_X$. If $N\lesssim 1$, then only the first excited state (the PIDM) is produced effectively, and we can consistently neglect all the higher KK modes in our calculations.

Given the bound on the energy scale of inflation, we can assume $N\lesssim 1$ for $m_X \lesssim 10^{-4}/10^{-5}m_p$ and $\gamma \gtrsim 0.01$. In this regime, we can expand the cross sections in powers of $T/m_X$ and retain the first few terms:
\begin{eqnarray}
\langle\sigma v\rangle_0&=&\frac{61 G^2 m_X^2}{1152 \pi } \left(1+\frac{243 }{61} \frac{T}{m_X}+\frac{3244 }{61 } \frac{T^2}{m_X^2}+O\left(\frac{T}{m_X}\right)^3\right)\,,\nonumber\\
\langle\sigma v\rangle_{1/2}&=&\langle\sigma v\rangle_{1}=\frac{175 G^2 m_X^2}{192 \pi } \left(1+3 \frac{ T}{m_X}+\frac{15243}{700} \frac{T^2}{m_X^2}+O\left(\frac{T}{m_X}\right)^3\right)\,.\nonumber\\
\end{eqnarray}
These are the (approximate) cross sections for the direct production of $KK_1$ by standard model particles, in complete analogy with what we have done for the lower spin cases.

\begin{figure}
\begin{center}
\includegraphics[width=0.7\textwidth]{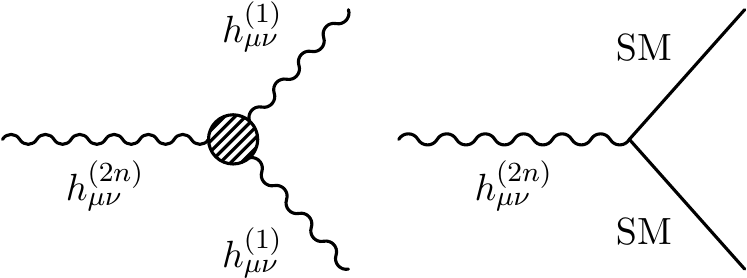}
\end{center}
\caption{\label{KK_decay}Decay of KK gravitons with even KK number into lighter KK modes (left) and SM particles (right).}
\end{figure}

In this scenario, however, we have another potentially competing process for the production of $KK_1$, namely the creation of a single even KK mode by standard model particles, which then decays to the PIDM. Even modes can couple directly to the standard model sector and thus be singly produced by the inverse decay process $SM+SM \rightarrow KK_{2n}$ of Fig.\,\ref{KK_invdecay}. The question then is wether these even modes decay predominantly back to SM particles or to the PIDM. The decay widths of KK gravitons to SM particles were computed in \cite{Han:1998sg}:
\begin{eqnarray}
\Gamma(KK \rightarrow S S)=\frac{1}{960 \pi}\frac{m^3}{m_p^2},\nonumber\\
\Gamma(KK \rightarrow f \bar{f})=\frac{3}{320 \pi}\frac{m^3}{m_p^2},\nonumber\\
\Gamma(KK \rightarrow V V)=\frac{N}{160 \pi}\frac{m^3}{m_p^2},\nonumber\\
\end{eqnarray} 
where $N$ is 1 for photons and 8 for gluons. The three partial decay widths describe decay to scalar, fermion and vector particles respectively. The total decay width to SM particles, which we denote by $\Gamma_{SM}$, will then just be a sum of the three, each counted with the right number of degrees of freedom. 

The decay width of a KK graviton to lighter modes can be computed simply from the triple graviton vertex, using the dimensional reduction prescriptions we described earlier.  In the regime $N \lesssim 1$ considered here, the dominant contribution will be given by the decay of the first few even modes, which are the lightest. The decay process $KK_2 \rightarrow 2 KK_1$ is classically forbidden because its phase space is exactly zero. However, radiative corrections to the mass of $KK_2$ could in principle open up the decay channel if they are big enough (and positive). 
Taking the mass of $KK_2$ to be $M$ and the one of $KK_1$ to be $m$, the decay rate is given by
\beq
\Gamma_{KK}=\frac{\sqrt{1-\frac{4 m^2}{M^2}} \left(M^2-4 m^2\right)^2 \left(90 m^8-48 m^6
   M^2+59 m^4 M^4+5 m^2 M^6+2 M^8\right)}{864 \pi m_p^2  m^8 M}\,.
\eeq

Next we have the decay $KK_4 \rightarrow 2 KK_1$, which vanishes at tree level due to KK number conservation in the bulk. However, the presence of the two branes explicitly breaks momentum conservation in the extra dimension, giving rise to a non-zero $KK_4$ contribution with loop-induced decay: KK number violating effects are localized on the orbifold fixed points and therefore only appear at loop level. 
All higher order decays have negligible contributions due to extreme Boltzmann suppression in the high mass limit. 
Given the number density of $KK_{2n}$, $n_{KK_{2n}}$, the additive contribution to the PIDM number density will then be $\delta n_X = 2n_{KK_{2n}} BR_{2n}$, where $BR_{2n}$ is the branching ratio of $KK_{2n} \rightarrow 2 KK_1$. As we said, for our purposes it is sufficient to consider only $n=1,2$.

Following \cite{Hall:2009bx}, the dimensionless abundance $X_{KK_{2n}}\equiv n_{KK_{2n}} a^3/T_{rh}^3$ for the process  $SM+SM \rightarrow KK_{2n}$ is
\beq
X_{KK_{2n}}=\frac{45 m_p \Gamma_{SM}}{(1.66) 4 \pi^4 M^2 g_{rh}^{3/2}} \int^{\infty}_{M/T_{rh}} K_1(x)x^3 dx,
\eeq
where $M=2n m$ is the mass of the $n=1,2$ mode and $g_{rh}\sim 100$ the effective number of standard model degrees of freedom at $T_{rh}$.  
If we call $X_D$ the PIDM abundance coming from direct production, the final abundance will be $X=X_D+2 X_{KK_2} BR_2 + 2 X_{KK_4} BR_4$.

Note that the presence of a process like $SM+SM \rightarrow KK_{2n}$ that leads to an additional contribution to the PIDM density is a qualitatively different feature of this model compared to the minimal PIDM scenario (with spin 0,1/2 and 1). The origin of this difference lies in the fact that for a maximally hidden dark sector, a dark matter particle of spin different than two can only communicate indirectly through gravity with the standard model sector, whereas a spin 2 particle, being graviton-like, can also couple directly to the standard model particles, without the need of a mediator. In this particular model the dark matter particle cannot couple directly to the standard model sector because of KK parity, but its heavier excitations can and this adds to the dark matter abundance when they decay to the stable mode.

\begin{figure}[h]
\begin{centering}
\includegraphics[width=12cm]{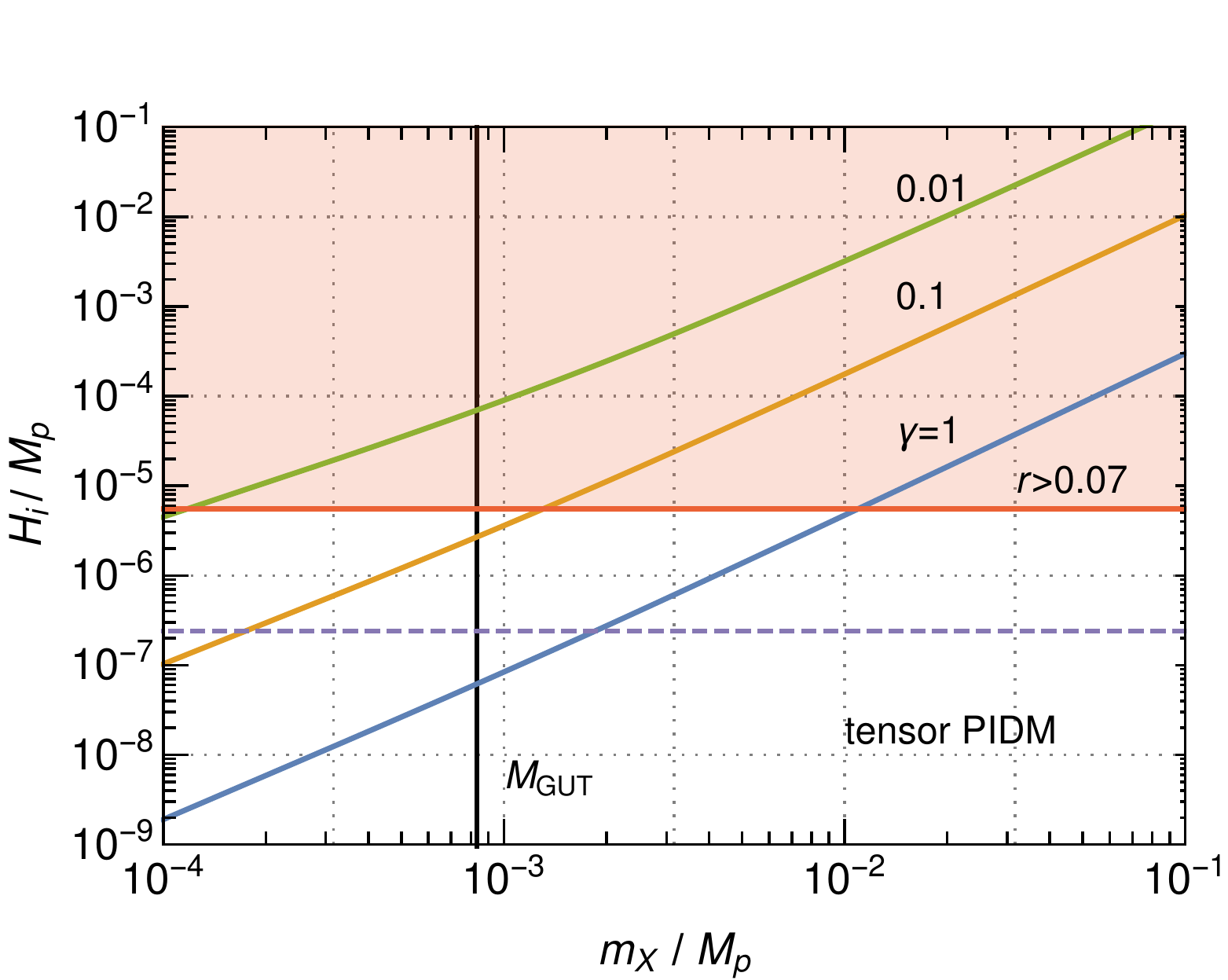}
\caption{Constraint plot for KK PIDM. The blue curve is for $\gamma=1$, orange for $\gamma=0.1$ and green for $\gamma=0.01$. 
The mass range is restricted to the interval $[10^{-4},0.1]$ where the single KK graviton approximation is valid. All values are given in units of $m_p$. }
\label{tensorplot}
\end{centering}
\end{figure}

Depending on the details of the model, both $BR_2$ and $BR_4$ can vary from 0 to 1, so that either process (direct production or inverse decay) can dominate. However, we checked numerically that the impact on the $H_i$ bound is very mild, due to the strong dependence of $X$ on $H_i$. The constraint plot is almost unaffected by a change in the branching ratios, proving the robustness of the predictions. 
The constraint plot is shown in Fig.\,\ref{tensorplot}.

The results are similar to the lower spin cases, the only difference being that the mass range for which we can trust our calculations is now significantly smaller. We checked that our results are stable when including higher order terms in the cross section formulae, which means that the single KK graviton approximation is valid. At lower masses the production of the other KK excitations become important and in principle one should take into account the full interacting theory of the KK tower with itself and the SM brane.

\subsection{Monodromy Inflation and PIDM}\label{mono}

Since the minimal PIDM with a GUT scale mass requires a high scale of inflation and efficient reheating, it is interesting to ask how the PIDM can be embedded into well-known models of high scale inflation. A setup that appears to overcome  problems of high scale inflation in string theory is monodromy inflation  \cite{Silverstein:2008sg,McAllister:2008hb,Kaloper:2008fb,Kaloper:2011jz,Kaloper:2016fbr,DAmico:2017cda}, which also leads to interesting possible signatures in terms of primordial non-Gaussianity  \cite{Hannestad:2009yx,Flauger:2010ja}.

In this subsection we briefly discuss how the PIDM can be incorporated in an effective description of monodromy inflation, where the effective 4-D monodromy potential is obtained from compactification of $11$-dimensional Supergravity (SUGRA) by the mixing of an axion-like field with a four-form from the effective four dimensional action \cite{Kaloper:2008fb}
\bea\label{Sinfl}
S_{\textrm{Infl}}
&=& \int d^4 x \sqrt{-g}\left[\frac{1}{16\pi^2} m_p^2 R -\frac{1}{2\cdot 4!}F_{\mu\nu\lambda\rho}F^{\mu\nu\lambda\rho}-\frac{1}{2}\partial_\mu\phi\partial^\mu\phi+\frac{\mu}{4!}\phi\epsilon^{\mu\nu\lambda\rho}F_{\mu\nu\lambda\rho}\right]\nonumber\\
& &+\frac{1}{6}\int d^4 x \sqrt{-g}\nabla_\mu\left[F^{\mu\nu\lambda\sigma}A_{\nu\lambda\sigma}-\mu\phi \frac{\epsilon^{\mu\nu\lambda\rho}}{\sqrt{-g}}A_{\nu\lambda\rho}\right]~.
\eea 
To incorporate the PIDM we add a massive scalar degree of freedom with a natural mass, $M$, of order the 11-dimensional Planck mass, $M_{11}$,
\beq\label{Spidm}
S_{\textrm{PIDM}} = -\frac{1}{2}\int d^4 x \sqrt{-g} \left[\partial_{\mu}\sigma\partial^\mu\sigma+M^2\sigma^2\right]~.
\eeq
And also add a mechanism of reheating via the coupling
\beq\label{Srh}
S_{\textrm{RH}} = \int d^4 x \sqrt{-g} \frac{\phi}{f_\phi}G_{\mu\nu}\tilde G^{\mu\nu}~,
\eeq
which parametrises the effective coupling of the inflaton to a Standard Model sector gauge field strength $G_{\mu\nu}$ (and its dual $\tilde G^{\mu\nu}$).  The effective model describing inflation, dark matter and reheating is
\beq
S_{\textrm{Eff}}=S_{\textrm{Infl}}+S_{\textrm{PIDM}}+ S_{\textrm{RH}} ~.
\eeq
We can for example think of $\sigma$ as a lightest stable Kaluza-Klein mode, and from the discussion of the previous section, the phenomenology is similar
when replacing it with a fermion or graviton mode, as long as it only couples gravitationally to the four-form and the inflaton. 

In the model of \cite{Kaloper:2008fb}, the four-form of eq.~(\ref{Sinfl}) describes a membrane moving in 11-dimensional SUGRA, and the background of the four-form breaks the shift symmetry providing a quadratic potential for the inflaton $\phi$, 
\beq
V_{\textrm{eff}}= \frac{1}{2}(q+\mu\phi)^2~,
\eeq
valid within the large field range $m_p\ll\phi \leq
 M_{11}^2/\mu$, with $\mu\sim\mathcal{O}(10^3)(M_{11}/m_p)^2 M_{11} \ll M_{11}$, such that large field inflation can take place. Since the amplitude of the observed density perturbations, $\delta\rho/\rho \sim 10^{-5}$ fixes the mass of the inflaton to be $\mu \sim 10^{13}$GeV, one obtains $M_{11}\sim M_{GUT} = 10^{16}$GeV. Thus, in this setup, we naturally obtain a PIDM with a mass $m_X\sim M_{11}$  of order the GUT scale.

In order to have an efficient production of the PIDM by graviton scattering, we need, as previously discussed, a high reheating temperature. In the model of  \cite{Kaloper:2008fb}, inflation ends when the inflaton background value, $q/\mu +\phi$,  drops below $M_{Pl}$ and the inflaton starts to oscillate in its potential. The axion-like inflaton naturally couples to the Standard Model gauge sector, with a coupling of the form (\ref{Srh}), with $\mu \lesssim f_\phi \lesssim M_{11}$. We can compute the reheating temperature by noting that the decay rate of the inflaton into the SM gauge field sector is
\beq
\Gamma=\frac{\mu^3}{8\pi f_\phi^2}~,
\eeq
which leads to a reheating temperature
\beq
\quad T_{\text{rh}}=\frac{\kappa_2}{(8\pi^2)^{1/4}}\frac{\mu^{3/2}m_p^{1/2}}{f_\phi}~.
\eeq
In terms of the parameter $\gamma = \sqrt{\Gamma/H_i}$, parametrising the efficiency of the reheating, we have
\beq
 \gamma=\frac{1}{\sqrt{8\pi}}\frac{\mu}{f_\phi}~.
\eeq
With the PIDM mass being set by $M_{11}$ of order the GUT scale, the right abundance of dark matter is obtained for $\gamma \gtrsim 0.1$, which implies that we have to require $f_\phi \sim \mu \sim 10^{13}$GeV. On the other hand, before ending this section, let us mention that it is also possible to lower the PIDM mass scale to be similar to the inflaton mass, $M\sim \mu$, while instead having the inflaton decay constant fixed to the GUT scale $f_\phi \sim M_{11} \sim 10^{16}$GeV. 

While we have demonstrated the the PIDM is natural in monodromy inflation from the point of view of effective field theory, it will be interesting to see a precise realization of the PIDM scenario in a full UV complete model of axion monodromy in string theory. This is, however, beyond the scope of the present paper.

\subsection{Higgs Inflation PIDMs}
Following the philosophy of minimality, it may also be interesting to ask if the PIDM can fit within scenarios of Higgs inflation \cite{Bezrukov:2007ep}. In these models inflation happens entirely within the standard model, modified only with a non-minimal coupling of the Higgs to the standard model. The Lagrangian is 
\beq
\mathcal{L}= \left( \frac{1}{16\pi^2}m_p^2 +\xi H^\dagger H\right) R+g^{\mu\nu} (D_\mu H)(D_\nu H)-\lambda \left((H^\dagger H)-\frac{v^2}{2} \right)^2~.
\eeq
Let us add to this setup a the action of the PIDM as in eq.~(\ref{Spidm}). 

The authors of \cite{Bezrukov:2014ipa} consider two examples, critical and non-critical Higgs inflation. In the original non-critical version of Higgs inflation, inflation happens on the plateau of the Starobinsky inflation form, which in the Einstein frame takes the form
\beq
V_{Star}  \approx \frac{\lambda m_p^4}{256\pi^4\xi^2}\left( 1+e^{-4\pi \phi / \sqrt{3} m_p}\right)^{-2}~,
\eeq
where $\phi$ is the canonically normalized Higgs field that acts as the inflaton. In this case $\xi \sim 10^4$ and the tensor-to-scalar ratio, $r = 16 \epsilon$, is related to the number of e-folds, $N_*$, when the CMB scales leave the horizon, through the relation $\epsilon = 3/(4N_*^2)$. In the non-critical case where reheating proceeds perturbatively, the reheating temperature was estimated to be $T_{rh} \sim 10^{14}$ GeV, which with $r=3\times10^{-3}$ implies $\gamma\sim 10^{-2}$, which is consistent with an effective Einstein frame PIDM mass of $m_X\sim 10^{-5}m_p$. For the relation to the Jordan frame see section (\ref{nmPIDM}). In this case, it may be interesting to explore a possible connection between the PIDM and leptogenesis, as the PIDM mass is around the mass of the heavy right handed neutrino masses. It has been argued in \cite{Bezrukov:2014ipa}, that in the non-critical case, Higgs inflation can proceed even if the standard model vacuum is metastable, since thermal corrections can restore the symmetry at the time of reheating in the non-critical case, and the inflaton will therefore be trapped in the electroweak vacuum. However, the instability may still be triggered during reheating \cite{Herranen:2014cua, Herranen:2015ima, Espinosa:2015qea}. In addition, there are serious concerns about the validity range of the effective theory used to describe non-critical Higgs inflation \cite{Burgess:2009ea, Barbon:2009ya, Burgess:2014lza} (see also \cite{Bezrukov:2010jz, Bezrukov:2017dyv} for
further discussions).

On the other hand, in the critical case one assumes that the top quark mass and the Higgs coupling are finely tuned against each other, such that second (potentially harmful) vacuum disappears and instead turns into an inflection point of the potential, that can be used for inflation. In this form  inflation no longer occurs at the plateau, and the relation 
$\epsilon \sim 1/N_*^2$ is broken. For a recent evaluation we refer to \cite{Iacobellis:2016eof}. A large tensor-to-scalar ratio is possible with $\xi \sim 10$ and a reheating temperature just below the GUT scale \cite{Hamada:2014iga,Bezrukov:2014bra}. In that case an effective Einstein frame PIDM with $m_X\sim M_{GUT}$  can fit within Higgs inflation.

\section{Additional Possible Signatures}\label{posssig}

\subsection{Nonperturbative Decay}

Throughout, we have posited the stability of the PIDM, nominally protected by a symmetry that is respected by all particle interactions.   However, quantum gravitational effects are expected to induce processes that violate global symmetries \cite{Banks:2010zn}. This means that the PIDM can in principle decay into SM particles through gravitational instanton interactions which are suppressed by the Euclidean action $S$. It is important to make sure that the stability of the PIDM is not spoiled by these nonperturbative effects, that will result in the production of ultra-high energy cosmic rays, as in \cite{Berezinsky:1997hy, Kuzmin:1997jua, Birkel:1998nx,Sarkar:2001se}. 

The possibility that the observed flux of ultra-high energy cosmic rays at $E\gtrsim 10^{19}$\,eV is dominantly produced by the decay of super-heavy dark matter has been excluded long ago based on the relative fraction of photons versus charged cosmic rays \cite{Abraham:2009qb, Abraham:2006ar}. Assuming instead that the observed flux
is of astrophysical origin, it is possible to put stringent upper limits on a potential exotic contribution due to dark matter decay.
Depending on the decay channel and its mass, these bounds can be translated in lower limits on the lifetime, that apply to the PIDM scenario. Observations from the AUGER observatory and the Telescope Array \cite{Abbasi:2015czo} place bounds for the lifetime for $q\bar{q}$ decay at $\gtrsim 10^{22}$\,yr \cite{Aloisio:2015lva} for masses in the range $M\sim 10^{13}-10^{16}$\,GeV. Additionally, bounds on an ultra-high energy neutrino flux analyzed in  \cite{Gondolo:1991rn, Esmaili:2012us, Aab:2015kma} constrain the lifetime of this decay to be $\gtrsim 10^{15}$\,yr for $M\sim 10^{16}$\,GeV. In addition, the invisible decay to relativistic particles (such as neutrinos) is constrained by Planck data \cite{Poulin:2016nat}, giving a
lower bound $\gtrsim 10^{11}$\,yr if PIDM provides the dominant contribution to the observed dark matter density.
In the future a signal at the energies indicative of PIDM decay could be potentially detected \emph{e.g.} 
at JEM-EUSO or at ARA \cite{Aloisio:2015lva}.

The PIDM decays, being mediated by instanton interactions, contain an additional parameter, namely the instanton action $S$. The experimental bounds can be translated to bounds on the value of this action.  
If nonperturbative quantum gravity effects are present, they can be described by additional operators in the effective Lagrangian that break the global symmetry and are suppressed by $e^{-S}$ \cite{Kallosh:1995hi}. The terms that will give the dominant contribution to the decay rate are the ones with the lowest power of $m_p$ in the denominator, i.e. the lowest dimensional symmetry breaking operators. The general form of these operators, for different spins, is listed here:
 \begin{eqnarray}
\text{spin 0}&:& \mathcal{L_{NP}}=g m_p X H^\dag H \nonumber\\[2pt]
\text{spin 1/2}&:& \mathcal{L_{NP}}=g \bar{X} H L \nonumber\\[2pt]
\text{spin 1}&:& \mathcal{L_{NP}}=g (\partial^\mu X^{\nu}-\partial^\nu X^\mu) B_{\mu\nu} \nonumber\\[2pt]
\text{spin 2}&:& \mathcal{L_{NP}}=\frac{g}{m_p}X^{\mu\nu} T^{SM}_{\mu\nu},
\end{eqnarray}
where $X$ is the PIDM field, $H (L)$ the Higgs (lepton) doublet, and $T^{SM}_{\mu\nu}$ is the energy-momentum tensor of the standard model fields.
The coupling constant $g \sim e^{-S}$ measures the strength of nonperturbative quantum gravity effects and is exponentially suppressed by the Euclidean action of the process. Note that the scalar operator is dimension 3, the fermion and vector are dimension 4, and the tensor operator is dimension 5. The decay lifetime of the PIDM for these types of operators can be estimated as 
\begin{eqnarray}
\text{dimension 3}: \tau_X=\alpha \frac{m_X}{m_p^2} e^{2S} \nonumber\\[2pt]
\text{dimension 4}: \tau_X=\alpha \frac{1}{m_X} e^{2S} \nonumber\\[2pt]
\text{dimension 5}: \tau_X=\alpha \frac{m^2_p}{m^3_X} e^{2S},
\end{eqnarray}
where $m_X$ is the mass of the PIDM and $\alpha$ is an $O(100)$ number. Due to the exponential dependence the lower limits on the instanton action $S$ are
rather insensitive to the details of the decay. For example, for GUT scale dark matter the requirement $\tau_X>10^{22}(10^{11})$\,yr translates into a lower
limit $S>85 (72)$ for the dimension 3 operator, $S> 77 (65)$ for dimension 4, and $S> 70 (58)$ for dimension 5.

How realistic are these values for the action? Following \cite{Kallosh:1995hi}, in a large class of models based on Einstein theory of gravity the action is fairly small, of order ${\cal O}(10^1)$. It is well-known, however, that the `wormhole' action is sensitive to the structure of spacetime on very small scales and to certain details of the quantum gravity theory that completes GR at high energies. Indeed, in the same paper, the authors showed that modifications of Einstein theory on very small length scales can lead to strong suppression of these nonperturbative effects, so that the dark matter particles may become essentially stable.

In order to assess the strength of these nonperturbative gravitational effects we need to compute the Euclidean action of a wormhole that can change the global charge, inducing a violation of the global symmetry. If we compute the action in Einstein gravity, this will be of order $S_E\sim m_p^2R^2$, where $R$ is the radius of the wormhole throat. If the description of spacetime is valid only up to a certain cutoff $\Lambda$, we should only include instanton contributions up to $\Lambda$. Then, the naive (though generically untrue) estimate of the instanton action is $S\sim m^2_p/\Lambda^2$, where it is implicitly assumed that the size of the wormhole is of the order of the cutoff length, $R \sim \Lambda^{-1}$. Therefore, if there are additional (lower) energy scales in our model (as in string theory and KK theories) above which classical gravity breaks down, gravitational instantons are strongly suppressed by factors of $e^{-m_p^2/\Lambda^2}$. 

If we take string theory as the quantum gravity theory, for example, violations of global symmetries are suppressed by the topological factor $e^{-8\pi^2/g^2}$, where $g$ is the string coupling. The string scale $M_s$ is given by $M^2_s=m^2_p g^2/(8\pi^2)$, so that we can rewrite this factor as $e^{-m^2_p/M^2_s}$.  In string models this effect is strong enough to completely erase any noticeable global symmetry breaking. A very large string coupling would be required in order to make this effect manifest.

\subsection{Decay through gravity portals}

In our model we imagine that the dark matter particle is stabilised by a mechanism that remains operational also in the presence of a curved background. In other words, we assume that the symmetry that stabilises the PIDM is valid also in a curved spacetime. It is interesting to see what happens when this assumption is dropped. The possibility of dark matter decay to standard model particles via interactions that are only present when the curvature is non-zero was explored in \cite{Cata:2016epa}. These interactions take the form of a non-minimal coupling between gravity and the dark matter field that breaks the global stabilising symmetry and induce the decay, and they are generically called ``gravity portals''.

The dominant contribution to the decay will be given by the lowest dimensional operator linear in the Ricci scalar $R$:
\beq\label{portal}
{\cal L}_{\xi}=-\xi R F(X)~,
\eeq
where $\xi$ is the dimensionless parameter controlling the strength of the non-minimal coupling and $F(X)$ a real function linear in the dark matter field $X$, which can carry any spin. After expressing the full theory in the Einstein frame, this term gives rise to dark matter decay into SM particles. 

For a scalar field, the function is simply $F(X)=MX$, where $M$ is a mass scale than can be taken to be of the order of the Planck scale. Then the non-minimal coupling term ${\cal L}_{\xi}=\xi M R \phi$ explicitly breaks the $Z_2$ stabilising symmetry under which the dark matter field $X$ is odd and the SM fields are even.  At tree level this term generates up to four-body decay. In general three-body decay is expected to dominate for low DM masses, while four-body decay is the dominant contribution for large DM masses. Taking into account all decay processes, as done in \cite{Cata:2016epa}, one can compute the total decay rate $\Gamma_X$ of DM to SM particles and compare this to the age of the universe to obtain a conservative bound on the DM mass and/or the coupling parameter $\xi$. Natural values of the non-minimal coupling parameter, $\xi \sim O(1)$ lead to lifetimes that are orders of magnitude shorter than the age of the universe in the whole range of masses that we consider: from $\sim 1$ GeV up to the GUT scale. This implies that, for scalar DM at least, the non-minimal coupling to gravity must be extremely suppressed, especially for large DM masses. A possible mechanism to realize this is to imagine that the scalar DM field is charged under a gauge symmetry that is spontaneously broken, effectively producing the gravity-portal term with a very small $\xi$.  

The situation improves somewhat if we consider non-scalar DM. We know for example that fermionic particles are protected against rapid decays by Lorentz symmetry.  In the case where $X$ is a fermionic singlet, it was proven in \cite{Cata:2016epa} that for $\xi \sim O(1)$, the lifetimes are consistent with observations if $m_X \lesssim 10^6$ GeV, which has some overlap with our mass range. For high masses, however, and in particular for natural values of the order of the GUT scale, we still need a strong suppression on $\xi$. 

Note that the non-minimal coupling in (\ref{portal}) is different from the one in (\ref{NMC}). In fact, the first directly couples the DM particle to the SM degrees of freedom, while the latter just describes an additional (besides the minimal one) direct coupling between DM (as well as the SM particles) and gravity. 

\section{Conclusions}

In this note, we have extended the original  scenario of Planckian Interacting Dark Matter laid out in  \cite{Garny:2015sjg}. In particular, we have extended the analysis to fermion, vector and spin 2 PIDM particles.  We find that the main conclusions we had obtained for the scalar case are robust, and that the broad details of the PIDM sector do not lead to qualitative changes.  Therefore, in order for the PIDM to be minimal, with mass close to the GUT scale, reheating must be both very efficient and occur at a very high scale.  This, in turn, implies the creation of a detectable level of tensor modes. Not seeing primordial tensor modes in the near future will therefore imply that dark matter either has non-gravitational interactions with the SM sector or that there is  a new mass scale of nature related to the dark matter mass.

We have also shown that it is possible to incorporate the PIDM scenario into existing theoretical frameworks, such as compactifications and monodromy, and have also commented on Higgs inflation.  Scenarios such as Horava-Witten compactifications, which do not have a known WIMP candidate \cite{Banks:1999dh}, may now be integrated into a realistic cosmology through this mechanism.  Though we prefer the PIDM mass to be around the GUT scale, incorporating it into a GUT theory immediately alters our mechanism, as generically, even heavy singlet states may interact with standard model particles through massive gauge bosons.
Additionally, GUT theories contain monopole states that will be overproduced during reheating, in conflict with experimental bounds.  If our motivation is taken seriously, and there is no additional physics beyond the standard model until very high energies, however, the absence of gauge coupling unification may point to the more minimal scenario discussed here, with gravitational interactions only.

Though the PIDM does not predict signals searched for in current dark matter direct or indirect detection experiments, there is a possibility that gravitationally mediated decays lead to production of ultra high energy cosmic rays, gamma rays and neutrinos that may potentially be detected in future observatories.  The very stability of this scenario, if guaranteed by a global symmetry, requires mildly large Euclidean action, which has implications for the types of quantum gravity theories one expects to ultimately find at high energies. On the other hand, if the stability is guaranteed by a discrete symmetry, the PIDM may also be absolutely stable.

The lack of conventional signals makes the minimal PIDM scenario falsifiable, in the sense that if any deviations from the pure cold dark matter framework are found, the minimal PIDM scenario will be ruled out.   
Any indication that dark matter is self interacting, warm, couples to standard model particles, or is produced in colliders, will eliminate the minimal PIDM scenario.  Within the minimal PIDM framework, all current dark matter anomalies, such as the missing satellite problem, the cusp-core problem, the too big to fail problem, the diversity problem, and the various signals seen from galactic environments, must ultimately be explained by conventional standard model processes, rather than exotic dark matter physics. We note that it is possible, however, to deviate from the minimal PIDM scenario either by adding new mass scales or additional self-interactions in the dark matter sector, while having dark matter interacting only gravitationally with the SM sector. In this non-minimal self-interacting PIDM scenario the PIDM can presumably be produced by the same mechanism as outlined in the present paper, but with additional non-trivial observational signatures. We leave it for future work to explore this possibility.
On the other hand, the minimal PIDM scenario with $M\gtrsim M_{GUT}$, that is motivated by naturalness arguments, requires a tensor-to-scalar ratio
$r\gtrsim 10^{-4}$, providing a benchmark value for future CMB polarization experiments.

\bigskip

{\bf \noindent Acknowledgements}

\smallskip
\noindent We  thank Nemanja Kaloper, Antonio Riotto, Subir Sarkar, and Masaki Yamada for interesting discussions. We thank Bohdan Grzadkowski for correcting a sign in our first version. MSS is supported by Villum Fonden grant 13384. CP3-Origins is partially funded by the Danish National Research Foundation, grant number DNRF90. 

\section*{Note added}

During the completion of this work the papers  \cite{Tang:2017hvq, Albornoz:2017yup,Dhital:2017yuu} appeared. The results of \cite{Tang:2017hvq} are related to our discussion in section 3, and in the cases of overlap between the two papers our results agree, except for a sign in the vector energy-momentum tensor affecting the vector amplitudes. In \cite{Albornoz:2017yup} the PIDM as a massive graviton is discussed, but the authors appear to only consider the light PIDM regime. In \cite{Dhital:2017yuu} the possible detection of PIDM cosmic ray decay signals are discussed.

\begin{appendix}

\section{Three-graviton vertex}\label{app:3g}

For the computation of the abundance of spin-2 PIDM within the framework of orbifold compactifications we
need three-graviton vertices involving two massive and one massless mode.
We follow the strategy outlined in \cite{Davoudiasl:2001uj}, and first consider the three-graviton vertex
in five dimensions, which has the same structure as in 4d \cite{Bjerrum-Bohr:2014lea}, but with all Lorentz indices replaced by five-dimensional
ones. By contracting the open indices with 4-dimensional polarization vectors or 4d-propagators, only the
four-dimensional part of the vertex is projected out. However, in addition, the vertex contains cross-products
of 5d momenta of the form $k_a^{(5)}\cdot k_b^{(5)}$. For the orbifold compactification, they can be
expressed in terms of four-momenta in the following way: the KK modes of the graviton $h_{\mu\nu}$ can be decomposed as (only even modes under KK contribute to $h_{\mu\nu}$ with $\mu,\nu$ in 4d)
\be
  \phi(X) \to \frac{1}{\pi R}\sum_n \phi^{(n)}(x)\cos\left(\frac{ny}{R}\right)
\ee
which yields the usual KK mass term ($G^{MN}=$diag$(1,-1,-1,-1,-1)$)
\be
 \int_0^{2\pi R}G^{MN}\partial_M \phi(X)\partial_N\phi(X) = \sum_n \left(\eta^{\mu\nu}\partial_\mu\phi^{(n)}(x)\partial_\nu\phi^{(n)}(x)-\frac{n^2}{R^2}(\phi^{(n)}(x))^2\right)
\ee
The 4d fields are decomposed in momentum space in the usual way $(\int_k=\int \frac{d^3k}{(2\pi)^32E_k})$,
\be
  \phi^{(n)}(x) = \int_k \left(e^{ikx}a_k^{(n)\dag} + e^{-ikx}a_k^{(n)} \right)
\ee
with $k^2=n^2/R^2$.

We are interested in contributions to the three-graviton vertex of the form ($d^5X=d^4x dy$)
\be
  V=\int d^5X \phi(X) G^{MN}\partial_M \phi(X)\partial_N \phi(X)
\ee
which gives $k_a^{(5)}\cdot k_b^{(5)}$ terms in 5d momentum space.
Let us consider terms with one zero-mode (usual graviton)
and two modes $n\not=0$. The relevant cross-product terms come about in contributions where the 5d-derivative acts 
on the $n\not=0$ modes,
\bea
  V &\supset& \int d^4x \phi^{(0)}(x) \int_0^{2\pi R} dy \left(\eta^{\mu\nu}\partial_\mu\phi^{(n)}(x)\partial_\nu\phi^{(n)}(x)\cos^2(ny/R) 
  - \frac{1}{R^2}(\phi^{(n)}(x))^2\sin^2(ny/R)\right) \nn\\
   &=& \pi R \int d^4x \phi^{(0)}(x) \left(\eta^{\mu\nu}\partial_\mu\phi^{(n)}(x)\partial_\nu\phi^{(n)}(x) 
  - \frac{1}{R^2}(\phi^{(n)}(x))^2\right)
\eea
Consider a $\phi^{(0)}(k)\to \phi^{(n)}(k_a)\phi^{(n)}(k_b)$ process,
\be
  \langle 0 | a_k^{(0)} \, V \, a_{k_a}^{(n)\dag}a_{k_b}^{(n)\dag} | 0 \rangle \propto \eta^{\mu\nu} (i k_a)_\mu (i k_b)_\nu -\frac{n^2}{R^2} = -(k_a \cdot k_b + \frac{n^2}{R^2})
\ee
When following along the lines of \cite{Davoudiasl:2001uj} to obtain the full three-graviton vertex this
gives the replacement rule
\be
  k_a^{(5)}\cdot k_b^{(5)} \to k_a \cdot k_b + \frac{n^2}{R^2}
\ee
for the dot-product of five-momenta in terms of the four-dimensional momenta and the KK mass.
Similarly, when considering the self-energy of the $n$th KK mode (one in- and one out-going momentum) one
finds $(k_a^{(5)})^2\to k_a^2-n^2/R^2=0$, in accordance with the naive expectation. Note that five-momentum
is in general not conserved due to the breaking of shift symmetry in the five-direction by the orbifolding.

For the decay $m=2n \to n+n$ of a heavier mode into two lighter modes, we also need the corresponding three-graviton vertex.
One finds analogously to above $k_a^{(5)}\cdot k^{(5)} \to k_a^{(5)}\cdot k_b^{(5)}$, $k^{(5)}\cdot k_b^{(5)} \to k_a^{(5)}\cdot k_b^{(5)}$,
and $(k^{(5)})^2\to 0$ where $k$ corresponds to the momentum of the KK mode $m=2n$. Note that one may \emph{not} replace
$(k_c^{(5)})^2$ by $ (k_a^{(5)}+ k_b^{(5)})^2$ here, due to the breaking of shift symmetry mentioned above, while four-momentum is conserved
in the usual way, $k =  k_a + k_b$.

\end{appendix}

\end{document}